\documentclass[iop]{emulateapj}

\usepackage{natbib} 
\usepackage{graphicx}        
\begin{document}
\title{Biosignature Gases in H$_2$-Dominated Atmospheres on Rocky Exoplanets}

\author{S. Seager\footnote{Dept. of Earth, Atmospheric and Planetary Sciences,
Massachusetts Institute of Technology, 77 Massachusetts Ave.,
Cambridge, MA, 02139.}$^{,}$\footnote{Dept. of
Physics, Massachusetts Institute of Technology, 77 Massachusetts Ave.,
Cambridge, MA, 02139}, {W. Bains$^{1,}$\footnote{Rufus Scientific}}, {R. Hu$^1$},} 

\begin{abstract}
Super Earth exoplanets are being discovered with increasing frequency
and some will be able to retain stable H$_2$-dominated atmospheres.
We study biosignature gases on exoplanets with thin H$_2$ atmospheres
and habitable surface temperatures, by using a model atmosphere with
photochemistry, and biomass estimate framework for evaluating the
plausibilty of a range of biosignature gas candidates. We find that
photochemically produced H atoms are the most abundant reactive
species in H$_2$ atmospheres. In atmospheres with high CO$_2$
  levels, atomic O is the major destructive species for some
  molecules. In sun-Earth-like UV radiation environments, H (and
  in some cases O) will rapidly destroy nearly all biosignature gases
of interest. The lower UV fluxes from UV quiet M stars would produce a
lower concentration of H (or O) for the same scenario, enabling
some biosignature gases to accumulate.  The favorability of low-UV
radiation environments to accumulation of detectable biosignature
gases in an H$_2$ atmosphere is closely analogous to the case of
oxidized atmospheres, where photochemically produced OH is the major
destructive species.  Most potential biosignature gases, such as DMS
and CH$_3$Cl are therefore more favorable in low UV, as compared to
solar-like UV, environments.  A few promising biosignature gas
candidates, including NH$_3$ and N$_2$O, are favorable even in
solar-like UV environments, as these gases are destroyed directly by
photolysis and not by H (or O). A more subtle finding is that
most gases produced by life that are fully hydrogenated forms of an
element, such as CH$_4$, H$_2$S, are not effective signs of life in an
H$_2$-rich atmosphere, because the dominant atmospheric chemistry will
generate such gases abiologically, through photochemistry or
geochemistry.  Suitable biosignature gases in H$_2$-rich atmospheres
for super Earth exoplanets transiting M stars could potentially be
detected in transmission spectra with the {\it James Webb Space
  Telescope}.
\end{abstract}

\section{Introduction}

The detection of exoplanet atmospheric biosignature gases by remote
sensing spectroscopy is usually taken as inevitable for the future of
exoplanets. This sentiment is being borne out with the discovery of
increasing numbers of smaller and lower mass planets each year.  In
addition, the natural evolution to development of larger and more
sophisticated telescopes (such as the {\it James Webb Space Telescope}
(JWST) slated for launch in 2018, \citet{gard2006}) and the giant 20-
to 40-meter class ground-based telescopes\footnote{The Extremely Large
Telescope (http://www.eso.org/public/teles-instr/e-elt.html), the
Giant Magellan Telescope (http://www.gmto.org/), and the Thirty Meter
Telescope (http://www.tmt.org/).} continues to fuel the concept
that the eventual detection and study of biosignature gases is a near
certainty.

The topic of biosignature gases, however, may remain a futuristic one
unless a number of extreme challenges can be overcome. The biggest
near-term challenge is to find a large enough pool of potentially
habitable exoplanets accessible for followup atmosphere
study\footnote{For example the all-sky space-based TESS mission
  (Transting Exoplanet Planet Survey Satellite, PI George Ricker) has
  been selected under NASA's Astrophysics Explorer Program for launch
  in 2017.}.  By potentially habitable we mean rocky planets with
surface liquid water, and not those with massive envelopes making any
planet surface too hot for complex molecules required for life. The
large pool of such planets is needed because there could be a large
difference in the numbers of seemingly potentially habitable planets
(based on their measured host star type, orbit, and mass or size and
inferred surface temperature) and those that are inhabited by life
that produces useful biosignature gases (which will be inferred from
measured atmospheric spectra). Useful biosignature gases means those
that can accumulate in the planet atmosphere, are spectroscopically
active, and are not overly contaminated by geophysical false
positives. A contemporary, related point on identifying a large enough
pool of planets is that even the fraction of small or low-mass planets
that are potentially habitable---that is with surface conditions
suitable for liquid water---is not yet known. The reason is 
that the factors controlling a given planet's surface temperatures are 
themselves not yet observed
or known, including the atmosphere mass (and surface pressure),
the atmospheric composition, and hence the concomitant
greenhouse gas potency \citep[see the review
  by][]{seag2013a}.

A second major challenge for the study of biosignature gases is the
capability of telescopes to robustly detect molecules in terrestrial
exoplanet atmospheres. This challenge is continously faced in today's
hot Jupiter atmosphere studies \citep[e.g.,][]{seag2010}, where many
atmospheric molecular detections based on data from the {\it Hubble
  Space Telescope} or the {\it Spitzer Space Telescope} remain
controversial \citep[see][and references therein]{demi2013}. For
transiting planets, the ability to identify and remove systematics to a
highly precise level while adding together numerous transit events
from different epochs is a necessity to reach the small signals of
terrestrial planet atmospheres.  For directly imaged planets, the
ability to reach down to Earth-sized planets in Earth-like orbits is
one of the most substantial technological challenges to ever face
astronomers.  While technology development is ongoing, there are as
yet no solid plans to launch a space telescope capable of directly
imaging terrestrial-size planets.

A third major challenge in the study of biosignature gases has to do
with the geological false positive signatures. These false positives
are gases that are produced geologically and emitted by volcanoes or
vents in the crust or ocean. Geochemistry has the same chemicals to
work with that life does, and therefore false positives are
inevitable. While early theoretical studies favored detection of redox
disequilibrium (such as O$_2$ and CH$_4$) that should not both exist
in an atmosphere in photochemical steady state, often one molecule of
the set is too weak spectroscopically for potential detection. The
conventionally adopted approach (at least in theoretical studies) is
therefore to identify a biosignature gas that is many orders of
magnitude out of thermodynamic equilibrium with the expected gas
composition of the atmosphere and to study the gas in the context of
the planet atmosphere environment via atmospheric spectra that cover a
wide wavelength range. A more likely outcome to the field of
biosignature gases will be to develop probabilistic assessment of the
likelihood a molecule in a given atmosphere can be attributed to life,
because spectroscopic data and the information for a complete
assessment of the planetary environment will be limited.

To increase the chances of detecting exoplanet atmospheric
biosignature gases we are motivated to widen the parameter space of
types of planets where biosignature gases can accumulate and should be
sought out observationally.  We here describe, for the first time to
our knowledge, the case for and against biosignature gases in
hydrogen-rich atmospheres. Some massive enough or cold enough super
Earths (loosely defined as planets with up to 10 Earth masses) will be
able to retain hydrogen in their atmospheres (see the discussion in
\S\ref{sec-Hescape}).  In general, planets are expected to outgas or
capture hydrogen from the nebula during planet formation. Here we are
concerned with super Earths with relatively thin hydrogen atmospheres
and not planets with massive atmospheres or envelopes (as in
mini-Neptunes) which will have surfaces too hot for liquid water
(Rogers and Seager, in prep.) or may not even have a surface.  A thin
hydrogen atmosphere does not add much to either the mass or the size
of the planet \citep{adam2008}, so that an H$_2$-rich atmosphere
itself does not aid in planet discovery or detection.

Super Earths with H$_2$-rich atmospheres are nonetheless in some ways
more favorable for detection and study than their terrestrial planet
counterparts with N$_2$- or CO$_2$-dominated atmospheres.  A more
massive planet than Earth (i.e., more likely to retain atmospheric
H$_2$ than Earth) is easier to discover than an Earth-mass planet via
the radial velocity technique.  A more massive planet than Earth is
also larger and so easier to discover or detect by the transit
technique than a lower-mass planet.  For example a 10~$M_\oplus$
planet of Earth-like composition would have a radius 1.75 times larger
than Earth \citep[e.g.,][]{seag2007}. The larger planet area is more
favorable for atmosphere study in reflected or thermally emitted
radiation than an Earth-size planet. For transit transmission spectra,
planets with H$_2$-rich atmospheres have a much larger signal compared
to H-poor atmospheres because of the larger scale height $H$, based on
the mean molecular weight $\mu$ \citep[e.g.,][]{seag2010},
\begin{equation}
\label{eq:scaleheight}
  H = \frac{kT}{\mu m_H g},
\end{equation}
where $k$ is Boltzmann's constant, $T$ is temperature, $m_H$ is the
mass of the hydrogen atom, and $g$ is the surface gravity. The point
is that when H$_2$ dominates the atmospheric composition over
terrestrial planet atmosphere gases CO$_2$ and N$_2$, the mean
molecular weight is $\sim$20 times smaller and hence the scale height
is $\sim$20 times larger. The observational imprint of an atmosphere
is usually taken as about $5H$.

\begin{figure*}[ht]
\begin{center}
\includegraphics[scale=.65]{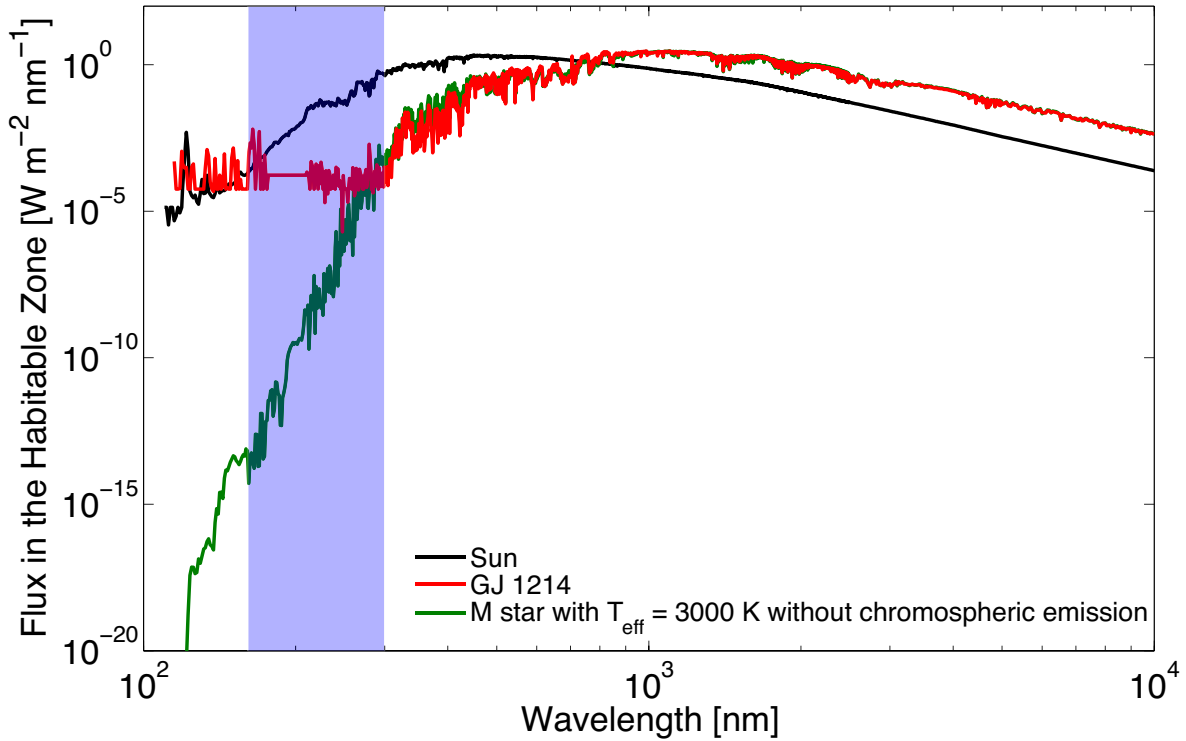}
\caption{Comparison of stellar fluxes. The radiative flux
    received by a planet in the habitable zone of a solar-like star, a
    weakly active M dwarf star (like GJ~1214), and a theoretically
    simulated quiet M dwarf star with an effective temperature of
    3000~K with no chromospheric emission. The flux is scaled so that
    the planet has a surface temperature of 290~K. The spectrum of the
    sun-like star is from the Air Mass Zero reference spectrum during
    a solar quiet period
    (http://rredc.nrel.gov/solar/spectra/am0/). The spectrum of
    GJ~1214 contains two parts: for wavelengths shorter than 300~nm we
    take the most recent HST measurement (France et al. 2013), for
    wavelengths longer than 300~nm we take the NextGen simulated
    spectrum for an M dwarf star having parameters closest to those of
    GJ~1214 (i.e., effective temperature of 3000~K, surface gravity $
    \log (g)=5.0$, and metallicity = 0.5). The spectrum with no
    chromospheric emission is also from the NextGen model (Allard et
    al. 1997). Under the common definition of weakly active, or
    relatively quiet M dwarf, the UV environment in its habitable zone
    can differ by more than six orders of magnitude.}
\label{fig:starflux}
\end{center}
\end{figure*}

Turning back to biosignature gases, they have been studied
theoretically as indicators of life on planets with oxidized
atmospheres for over half a century, beginning with \citet{lede1965}
and \citet{love1965}.  One highlight from the last decade is the
realization that low UV radiation environments compared to solar lead
to a much higher concentration of biosignature gases, as studied for
Earth-like planet atmospheres. This is because the stellar UV creates
the radical OH (in some cases O) which destroys many gases in the
atmosphere and thus reduces the gas lifetime. A low UV radiation
environment is taken to be that of a planet orbiting a UV quiet M dwarf
star  (see Figure~\ref{fig:starflux} and discussion in 
\S\ref{sec-Mstarradiation}).

A second highlight in biosignature gas research in the last decade is
the theoretical exploration of potential biosignature gases beyond the
conventionally considered dominant Earth or early Earth based ones of
O$_2$, O$_3$, N$_2$O, and CH$_4$. The variations studied include
dimethylsulfide \citep[DMS;][]{pilc2003}, methyl chloride
\citep[CH$_3$Cl;][]{segu2005}, and other sulfur compounds including
CS$_2$ and COS \citep{doma2011}. See \citet{seag2012} for a review,
\citet{seag2013b} for a biosignature gas classification scheme, and
\citet{seag2013b} for a biomass model estimate intended as a
plausability check to consider biosignature gas surface fluxes
different from Earth values.

We begin with a description of our atmosphere and biomass estimate
model in \S2. We present general results in \S3 and specific results
for a number of potential and unlikely biosignature gases in \S4. A
discussion in \S5 is followed by a summary and conclusion in \S6.

\section{Model}

The model goal is to computationally generate atmospheric spectra for
exoplanets with H$_2$-rich atmospheres with biosignature gases. The
model consistes of a photochemistry code which takes biosignature
gases as surface fluxes, an approximate temperature profile
calculation, and a line-by-line spectral calculation \citep{seag2013b,
  hu2012}.  The model also uses a biomass model estimate, to check
whether or not a biosignature gas could be the result of a plausible
surface ecology.

\subsection{Model Atmosphere}
{\it Photochemistry Model}
\label{sec-photochemmodel}
The focus on chemistry is critical for biosignature gases, because
sinks control a biosignature gas lifetime and hence the gas'
  potential to accumulate in the planetary atmosphere. A model for
  atmospheric chemistry is required to connect the amount of
  biosignature gas in the atmosphere (as required for detection) to
  the biosignature source flux at the planetary surface.

Our photochemical model is presented in \citet{hu2012}. The
photochemical code computes the steady-state chemical composition of
an exoplanetary atmosphere. The system can be described by a set of
time-dependent continuity equations, one equation for species at each
altitude. Each equation describes: chemical production; chemical loss;
eddy diffusion and molecular diffusion (contributing to production or
loss); sedimentation (for aerosols only); emission and dry deposition
at the lower boundary; and diffusion-limited atmospheric escape for
light species at the upper boundary. The code includes 111 species,
824 chemical reactions, and 71 photochemical reactions. 
Starting from an initial state, the system is numerically evolved to
the steady state in which the number densities no longer
change. 

The generic model computes chemical and photochemical
reactions among all O, H, N, C, S species, and formation of sulfur and
sulfate aerosols. The numerical code is designed to have the
flexibility of choosing a subset of species and reactions in the
computation. The code therefore has the ability to treat both oxidized
and reduced conditions, by allowing selection of ``fast species''.  For
the chemical and photochemical reactions, we use the most up-to-date
reaction rates data from both the NIST database
(http://kinetics.nist.gov) and the JPL publication (Sander et
al. 2011). Ultraviolet and visible radiation in the atmosphere is
computed by the $\delta$-Eddington two-stream method with molecular
absorption, Rayleigh scattering and aerosol Mie scattering
contributing to the optical depth. The model was developed
from the ground-up and has been tested and validated by
reproducing the atmospheric composition of
Earth and Mars \citep{hu2012,huthesis2013}. 

For biosignature gases that are minor chemical perturbers in the
atmosphere, the biosignature lifetime can be estimated based on the
abundance of the major chemical sink.  For this paper, the values
  of NH$_3$ and N$_2$O surface source fluxes are calculated from the
  full photochemistry model, whereas the calculations of surface
  source fluxes for other biosignature gases are simplified
  estimates.  One more point to note is that photochemistry is
relevant high in the atmosphere typically above mbar levels to which
stellar UV radiation can penetrate from above.
 
{\it Temperature-Pressure Profile} 
The precise temperature-pressure
structure of the atmosphere is less important than photochemistry for
a first-order description of biosignatures in H$_2$-rich atmospheres. The
reason is that most biosignature gases of interest have sources and
sinks not signficantly affected by minor deviations in the temperature
pressure profile. Morever the biosignature gases themselves are
secondary players in governing the heating structure of the
atmosphere. 

We therefore justify using the photochemistry model in a stand-alone
mode, with a pre-calculated temperature-pressure
profile.  The calculated temperature-pressure profile is approximate
and is one which assumes a surface temperature (e.g., 290~K), an
appropriate adiabatic lapse rate for H$_2$-rich compositions, and a
constant temperature above the convective layer of the atmosphere.
Such assumed temperature profiles are consistent with greenhouse
warming in the troposphere and lack of ultraviolet absorber in the
stratosphere. The semi-major axis of the planet is then derived based
on the assumed temperature profile by balancing the energy flux of
incoming stellar radiation and outgoing planetary thermal
emission. The details of this procedure are described in
\citet{hu2012}.

{\it Synthetic Spectra} To generate exoplanet transmission and thermal
emission spectra, we use a line-by-line radiative transfer code
\citep{seag2000b, madh2009, hu2012}. Opacities are based on molecular
absorption with cross sections computed based on data from the HITRAN
2008 database \citep{roth2009}, molecular collision-induced absorption
when necessary \citep[e.g.,][]{bory2002}, Rayleigh scattering, and
aerosol extinction computed based on Mie theory. The atmospheric
transmission is computed for each wavelength by integrating the
optical depth along the limb path \citep[as outlined in, e.g.,][]
{seag2000a, mill2009}. The planetary thermal emission is computed by
integrating the radiative transfer equation without scattering for
each wavelength \citep[e.g.,][]{seag2010}.

We consider clouds in the emergent spectra for thermal emission by
considering 50\% cloud coverage by averaging a cloudy and cloud-free
spectra. We omit clouds for the transmission spectra model because
the clouds are at low altitudes whereas the spectral features form
at high altitudes.

\subsection{Biomass Model Estimates}
\label{sec-biomassmodels}
A biomass model estimate has been developed by \citet{seag2013b} that
ties biomass surface density to a given biosignature gas surface
source flux. The motivating rationale is that with a biomass estimate,
biosignature gas source fluxes can be free parameters in model
predictions, by giving a physical plausibility check in terms of
reasonable biomass. The approach aims to enable consideration of a
wide variety of both gas species and their atmospheric concentration
to be considered in biosignature model predictions.  The biomass model
estimates are valid to one or two orders of magnitude. We provide
  a summary of the biomass model here with the full details available
  in \citet{seag2013b}.

The biomass model is used in the following algorithm.  First we
calculate the amount of biosignature gas required to be present at
``detectable'' levels in an exoplanet atmosphere from a theoretical
spectrum (we define a detection metric in
\S\ref{sec-detectionmetric}).  Second, we determine the gas source
flux necessary to produce the atmospheric biosignature gas in the
required atmospheric concentration.  The biosignature gas atmospheric
concentration is a function not only of the gas surface source flux,
but also of other atmospheric and surface sources and sinks. Third, we
estimate the biomass that could produce the necessary biosignature gas
source flux. Fourth, we consider whether the estimated biomass surface
density is physically plausible, by comparison to maximum terrestrial
biomass surface density values and total plausible surface biofluxes.

Based on life on Earth, a summary overview is that a biomass surface
density of 10~g~m$^{-2}$ is sensible, 100~g~m$^{-2}$ is plausible, and
5000~g~m$^{-2}$ is possible. In real situations, the total biomass is
nearly always limited by energy, bulk nutrients (carbon, nitrogen),
trace nutrients (iron, etc.) or all three. Regarding global surface
biofluxes we provide values and references where needed in our results
and discussion.

The biomass model estimates are tied to the type of biosignature gas
and so we briefly summarize our biosignature classification scheme
before discussion each biomass model estimate.

{\it Type I biosignature gases} are generated as byproduct gases from
microbial energy extraction. 
For example, on Earth many microbes extract energy from chemical energy
gradients using the abundant atmospheric O$_2$ for aerobic
oxidation,
\begin{equation}
\label{eq:TypeIexample}
{\rm X + O_2 \rightarrow oxidized \hspace{0.05in} X  }.
\end{equation}
For example: H$_2$O is generated from H$_2$; CO$_2$ from organics;
SO$_2$ or SO$_4^{2-}$ from H$_2$S; rust from iron sulfide (FeS);
NO$_2^{-}$ and NO$_3^{-}$ from NH$_3$; etc.

On an exoplanet with an H$_2$-rich atmosphere, the abundant
reductant would  now atmospheric H$_2$ such that
\begin{equation}
{\rm H_2 + X \rightarrow  reduced \hspace{0.05in} X }.
\end{equation}
The oxidant must come from the interior.

In other words, for chemical potential energy gradients to exist on a
planet with an H$_2$-rich atmosphere, the planetary crust must (in part)
be oxidized in order to enable a redox couple with the reduced
atmosphere. The byproduct is always a reduced gas, because in a
reducing environment H$_2$-rich compounds are the available reductants.
To be more specific, oxidants would include gases such as
CO$_2$ and SO$_2$.

The Type I biosignature gas biomass model is based on thermodynamics
and is derived from conservation of energy and discussed in detail in
\citet{seag2013b}. The biomass model estimate is
\begin{equation}
\label{eq:TypeIBiomassModel}
\Sigma_B \simeq \Delta G \left[ \frac{F_{\rm source}}{ P_{m_e}} \right].
\end{equation}
Here $\Sigma_B$ is the biomass surface density in g~m$^{-2}$, $\Delta
G$ is the Gibbs Free energy of the chemical redox reaction from which
energy is extracted (e.g., equation~(\ref{eq:TypeIexample})).  
  $\Delta G$ depends on the standard free energy of reaction ($\Delta G_0$),
  and the concentration of the reactants and products. Reactant and
  product concentrations can include ocean pH (concentration of H$^+$) in
  reactions that generate or consume protons. $\Delta G_0$ values are taken
  from \citet{amen2001}.

The term $P_{m_{e}}$ is an empirically determined microbial
maintenance energy consumption rate, that is, the minimum amount of
energy an organism needs per unit time to survive in an active state
(i.e., a state in which the organism is ready to grow).  An empirical
relation has been identified by \citet{tijh1993} that follows an
Arrhenius law
\begin{equation}
\label{eq:Pme}
P_{m_e} = A \exp\left[ \frac{-E_A}{RT} \right].
\end{equation}
Here $E_A = 6.94 \times 10^4$~kJ~mol$^{-1}$ is the activation energy,
$R$ = 8.314~kJ~mol$^{-1}$ K$^{-1}$ is the universal gas constant, and
$T$ in units of K is the temperature. The constant $A$ is $4.3 \times
10^7$~kJ~g$^{-1}$~s$^{-1}$ for aerobic growth and $2.5 \times
10^7$~kJ~g$^{-1}$~s$^{-1}$ for anaerobic growth \citep{tijh1993}.
Here per g refers to per g of wet weight of the organism.  $P_{m_e}$
is in units of kJ~g$^{-1}$~s$^{-1}$.

The free parameter in this biomass model estimate
(equation~(\ref{eq:TypeIBiomassModel})) is the biosignature gas source
flux $F_{\rm source}$ (in units of mole m$^{-2}$ s$^{-1}$).  $F_{\rm
  source}$ is the flux of the metabolic byproduct and is also the
surface bioflux required to generate a given biosignature gas
concentration in the atmosphere.

{\it Type II biosignature gases} are byproduct gases produced by the
metabolic reactions for biomass building, and require energy. On Earth
these are reactions that capture environmental carbon (and to a lesser
extent other elements) in biomass. Type II biosignature reactions are
energy-consuming, and on Earth the energy comes from sunlight via
photosynthesis.

There is no useful biomass model for Type II biosignature
gases because once the biomass is built a Type II biosignature gas is
no longer generated.

{\it Type III biosignature gases} 
are produced by life but not as byproducts of their central chemical
functions. Type III biosignature gases appear to be special to
particular species or groups of organisms, and require energy for
their production.  Because the chemical nature and amount released for
Type III biosignature gases are not linked to the local chemistry and
thermodynamics, the Type III biosignature gas biomass model is an
estimate based on lab culture production rates. 

We estimate the biomass surface density by taking the biosignature gas
source flux $F_{\rm source}$ (in units of mole m$^{-2}$ s$^{-1}$)
divided by the mean gas production rate in the lab $R_{\rm lab}$ (in
units of mole~g$^{-1}$~s$^{-1}$), 
\begin{equation}
\label{eq:TypeIIIBiomassModel}
\Sigma_B \simeq \frac{F_{\rm source}}{R_{\rm lab}}.
\end{equation}
We take the maximum observed for the Type III $R_{\rm
  lab}$ rates $F_{\rm field}$ values from different studies
\citep{seag2013b}.  The caveat of the Type III biomass estimate
explicitly assumes that the range of $R$ for life on exoplanets is
similar to that for life in Earth's lab environment. Nonetheless we
have showed based on Earth's values that the Type III biomass model is
valid to one or two orders of magnitude. The goal, again, is to use
the biomass estimate to argue for or against plausibilty of a
biosignature gas based on Earth's biomass surface density values and
not for any prediction of quantitative values. 

{\it Bioindicators} are defined as the end product of chemical
reactions of a biosignature gas.

{\it Model caveats} are related to the order
  of magnitude nature of the biomass estimates, the possible
  terracentricity of the biomass model estimates, and the lack of
  ecosystem context (see \citet{seag2013b} \S6.1, 6.2, 6.3 for a detailed
  discussion). Here we provide a summary overview.  

The order of magnitude nature of the Type I biomass estimate derives
from the dependency of the estimate on $P_{m_e}$, itself
very sensitive to temperature. The possible terracentricity of our
estimates is related to use of $P_{m_e}$, which is derived from
observations of terrestrial microorganisms, but we argue the
dependency is largely on thermodynamics \citep{seag2013b}.  The order
of magnitude nature of the Type III biosignatures derives the reliance
on laboratory rates for microbial production rates, and this is also
possibly a terracentricity issue.

The lack of ecosystem context is a major limitation for the biomass
estimate. An ecosystem contains not only the producers (i.e., the
biomass estimate derived ultimately from the bioflux $F_{\rm source}$)
but also consumers, whereas the biomass model estimate considers only
the producers. In this sense the biomass estimate in
equation~(\ref{eq:TypeIBiomassModel}) is a minimum. We can fairly say
that in the case of a very small or very large biomass estimate, the
assessment of biosignature gas plausibility is valid: a small biomass
estimate gives room for consumers even as a minimum biomass and a
large biomass estimate as a minimum will remain large regardless of
the consumers. For the intermediate case where a large but not unreasonable
biomass is needed to generate a detectable biosignature, the decision
on whether the gas is a plausible biosignature is more complicated,
and will depend on the planetary context: geochemistry, surface conditions,
atmospheric composition and other factors.  

Again, we do not argue the biosignature biomass model estimates are an
accurate prediction of an extraterrestrial ecology, rather we
emphasize the goal of the biomass model estimates is the order of
magnitude nature for a first order asssessment of the plausability of
a given biosignature gas candidate.

\subsection{Detection Metric}

\label{sec-detectionmetric}

We now describe our metric for a ``detection'' that leads to a
required biosignature gas concentration.  For now, the detection has
to be a theoretical exercise using synthetic data. We determine the
required biosignature gas concentration based on a spectral feature
detection with a SNR=10.  Specifically, we describe the SNR of the
spectral feature as the difference between the flux in the absorption
feature and the flux in the surrounding continuum (on either side of
the feature) taking into account the uncertainties on the data,
\begin{equation}
{\rm SNR} = \frac{\left| F_{out} - F_{in} \right|}{\sqrt{\sigma^2_{F_{out}} +
  \sigma^2_{F_{in}}}},
\end{equation}
where $F_{in} \pm \sigma_{F_{in}}$ is the flux density inside the
absorption feature and $F_{out} \pm \sigma_{F_{out}}$ is the flux
density in the surrounding continuum, and $\sigma$ is the uncertainty
on the measurement.

The uncertainties of the in-feature flux and continuum flux are
calculated for limiting scenarios. For thermal emission we consider a
futuristic space telescope able to block out the light of the host
star.  The uncertainties of the in-feature flux and continuum flux are
calculated for a limiting scenario: an 1.75 times Earth-sized planet
orbiting a star\footnote{Assuming perfect removal of starlight.} at 10
pc observed (via direct imaging) with a 6 m-diameter telescope mirror
operating within 50\% of the shot noise limit and a quantum efficiency
of 20\%. The integration time is assumed to be 20 hours.  We note that
collecting area, observational integration time, and source distance
are interchangeable depending on the time-dependent observational
systematics.  This telecope scenario is based on a TPF-I type
telescope \citep{laws2008}.

For transit transmission spectra, we use the same equation as above,
but with the denominator replaced by the noise in the stellar flux
($F_*$), as in $\sqrt{(4\sigma_F^2)}$, because transmission
observations measure the difference between the in-transit stellar
flux and out-of-transit stellar flux. For transmission spectra we
consider a 6.5-m space telescope, having quantum efficiency of 0.25
observing with 50\% photon noise limit, with integration time of 60
hours in-transit and 60 hours out-of-transit (assuming observing of
multiple transits).  Again, we note that collecting area,
observational integration time, and source distance are
interchangeable depending on the time-dependent observational
systematics. This scenario is based on the JWST.

\section{Photochemistry Results: H is the Dominant Photochemically-Produced Reactive Species in
H$_2$-Rich Atmospheres}

\begin{figure}[ht]
\centering
\includegraphics[scale=.45]{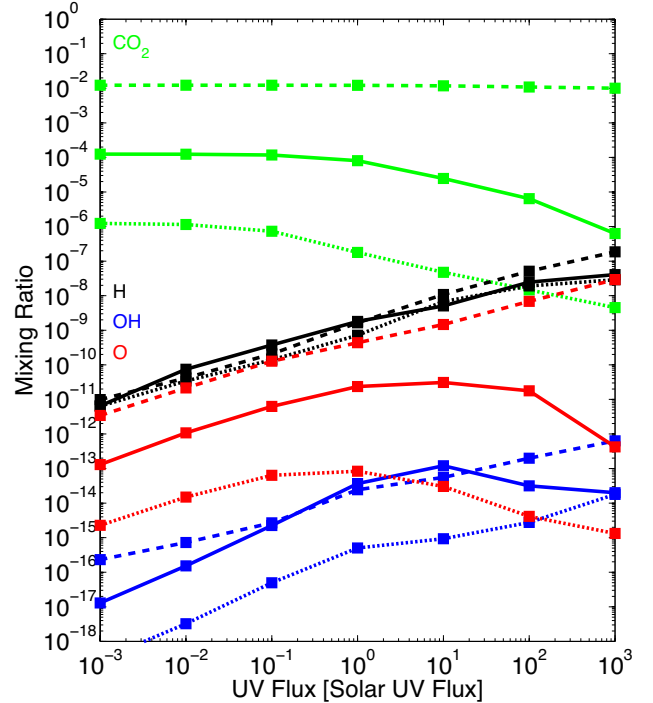}
\caption{Mixing ratio dependence of the reactive species (H, OH, and
  O) on UV flux in a H$_2$-rich atmosphere with some CO$_2$. Shown are
  different CO$_2$ levels.  The curves correspond to a CO$_2$ surface
  emission flux of Earth's volcanic emission rate ($3 \times 10^{15}$
  m$^{2}$ s$^{-1}$; solid lines), CO$_2$ emission rate 100 times
  higher than Earth's rate (dashed lines) and CO$_2$ emission rate
  100 times lower than Earth's rate (dotted lines). 
The planet has 10~M$_{\oplus}$ and 1.75~R$_{\oplus}$ and
  is in a 1.6-AU orbit of a sun-like star (with UV adjusted). The
  fiducial atmosphere is 90\% H$_2$ and 10\% N$_2$ by volume, in a
  1~bar atmosphere with a 290~K surface temperature.  The main
  point is that the H concentration does not depend on the amount of
  CO$_2$ in the atmosphere, whereas the amount of O is critically
  controlled by the level of CO$_2$ in the atmosphere. Compared to H,
  OH is always a minor constituent in the atmosphere (by a few orders
  of magnitude). As the UV flux increases, more of the destructive,
  reactive species are generated.}
\label{fig:mixingratiosvaryingCO2}
\end{figure}

In an H$_2$-rich terrestrial exoplanet atmosphere, atomic H is the
largest sink for most atmospheric molecules including biosignature
gases.  This is in contrast to oxidizing atmospheres (atmospheres with
substantial O$_2$ or CO$_2$ and H$_2$O and without H$_2$) where the OH
radical (and in some cases O) plays the role of the dominant sink.  We
note that for H$_2$-rich atmospheres with high CO$_2$ levels, atomic O
will be abundant (Figure~\ref{fig:mixingratiosvaryingCO2}) and for
some molecules will dominate the removal chemistry.

To explain the high H concentration we review the production of H, OH and O in
H$_2$-rich atmospheres. To qualitatively outline the main points we
use a simplified description of the main chemical pathways. This
discussion serves for illustration only, and is later backed up with a
more detailed computational photochemistry model.

To derive atmospheric concentrations of a species [A] we take
photochemical equilibrium,
\begin{equation}
\frac{d[A]}{dt} = P - [A]L = 0,
\end{equation}
\begin{equation}
\label{eq:pal}
[A] = \frac{P}{L}.
\end{equation}
where ${\rm[A]}$ is the mixing ratio of species ${\rm A}$, and $P$ and
$L$ are the production and loss rates respectively of species ${\rm
  A}$.  Below, the $K$ are reaction constants and $J$ is the
photodestruction rate associated with a stated reaction.

We consider an H$_2$ atmosphere with some H$_2$O.
\begin{equation}
\label{eq:H1}
{\rm H_2O + h \nu \rightarrow H + OH} \hspace{0.5cm} J,
\end{equation}
\begin{equation}
\label{eq:H2}
{\rm OH + H_2  \rightarrow H_2O + H} \hspace{0.5cm} K,
\end{equation}
\begin{equation}
\label{eq:H3}
{\rm H + H + M \rightarrow H_2 + M} \hspace{0.5cm} K_m.
\end{equation}
Combining the above two equations we have
\begin{equation}
\label{eq:Hconcentration}
{\rm [H]} = \sqrt{\frac{K {\rm [OH][H_2]}}{K_m {\rm [M]}}} 
= \sqrt{\frac {J {\rm [H_2O]}}{K_m}},
\end{equation}
and
\begin{equation}
{\rm [OH]} =  \frac{J {\rm [H_2O]}}{K {\rm [H]}}.
\end{equation}

The simplified atmosphere reveals a number of relevant points. The
first major point is that the role of OH in forming H from H$_2$
(equation~(\ref{eq:H2})) illustrates the importance of water
vapor. Water is needed to form H in the first place, in this
case. 

The second major point is that the reason H can accumulate to high
concentrations is because the H + H + M reaction rate that removes the
H atoms is relatively slow. The rates are
\begin{equation}
K = 2.8 \times 10^{-18} \exp(-1800.0/T) \hspace{0.1in}{\rm [m^3 s^{-1}]},
\end{equation}
\begin{equation}
K_m = 6.64 \times 10^{-39} (T/298.0)^{-1} \times N
\hspace{0.1in}{\rm [m^3 s^{-1}]},
\end{equation}
\begin{equation}
J \simeq 10^{-6} \hspace{0.1in}{\rm [s^{-1}]},
\end{equation}
where $N$ is the number density of species $M$ in units of 
molecules~m$^{-3}$. The rates are from \citet{sand2011}.

The third major point is that in the H$_2$-rich atmosphere the OH
concentration is low, because [OH] reacts with H$_2$ to recombine to
H$_2$O.

H is produced by photodissociation of water vapor and not
predominantly by the direct photodissociation of H$_2$.  The reason is
that the photons with high enough energy \citep[$\lambda <
  85$~nm;][]{ment1970} to photodissociate H$_2$ are not available.
The high-energy photons that could dissociate H$_2$ directly are
absorbed at the pressure levels of nanobars by H$_2$ itself, and the
photons that could penetrate down to pressure levels relevant to
observations (0.1~mbar to 1~bar) are those that can dissociate water.
Photodissociation of H$_2$O, in comparison, is caused by photons of
lower energy \citep[$\lambda \leq 240$~nm;][]{bank1973}, photons which
can penetrate more deeply in the atmosphere than the ones that
photodissociate H$_2$. We note that like other photochemical products,
H is formed primarily above the mbar level, before all of the
photodissociating stellar photons are absorbed.

The H concentration is dependent on the stellar UV levels and presence of
H$_2$O. Low-UV environments are favorable for biosignature build up,
since the initial photolysis that starts the OH formation chain will
be weaker. A similar situation is described for
oxidized atmospheres in \citet{segu2005}.

 We must beware that for some molecules, in some situations,
  atomic O will be the dominant destructive species. There is no
  simple model (as in the above equations) but with our full
  simulation we find in atmospheres with high CO$_2$ atomic O will
  abundant (see Figure~\ref{fig:mixingratiosvaryingCO2}). The key
  point is that reaction rates with O are faster than reaction rates
  with H for some molecules (see Table~\ref{tab-reactionrates}).

\begin{table*}
\begin{center}
\begin{tabular}{|l|lll|lll|}
\tableline
\tableline
Reaction & $A$ & $n$ & $E$ & $T=270$ K & $T=370$ K & $T=470$ K \\ 
\tableline
{DMS + H $\rightarrow$ CH$_3$SH + CH$_3$ }  & $4.81\times10^{-18}$ & 1.70 & 9.00   & $2.63\times10^{-24}$ & $2.11\times10^{-22}$ & $2.91\times10^{-21}$ \\
{CH$_3$Cl + H $\rightarrow$ CH$_3$ + HCl }  & $1.83\times10^{-17}$ & 0 & 19.29  & $1.97\times10^{-24}$ & $1.46\times10^{-22}$ & $1.92\times10^{-21}$ \\
{CH$_3$Br + H $\rightarrow$ CH$_3$ + HBr }  & $8.49\times10^{-17}$ & 0 & 24.44  & $1.59\times10^{-21}$ & $3.01\times10^{-20}$ & $1.63\times10^{-19}$ \\
{CH$_3$I + H $\rightarrow$ CH$_3$ + HI }  & $2.74\times10^{-17}$ & 1.66 & 2.49  & $7.67\times10^{-18}$ & $1.75\times10^{-17}$ & $3.09\times10^{-17}$  \\
\hline
{DMS + OH $\rightarrow$ CH$_3$SCH$_2$ + H$_2$O }  & $1.13\times10^{-17}$ & 0 & 2.10   & $4.43\times10^{-18}$ & $5.71\times10^{-18}$ & $6.60\times10^{-18}$ \\
{CH$_3$Cl + OH $\rightarrow$ CH$_2$Cl + H$_2$O} & $1.40\times10^{-18}$ & 1.60 & 8.65  & $2.54\times10^{-20}$ & $1.89\times10^{-19}$ & $3.17\times10^{-19}$ \\
{CH$_3$Br + OH $\rightarrow$ CH$_2$Br + H$_2$O }  & $2.08\times10^{-19}$ &1.30 & 4.16  & $2.87\times10^{-20}$ & $7.13\times10^{-20}$ & $1.30\times10^{-19}$ \\
{CH$_3$I + OH $\rightarrow$ CH$_2$I + H$_2$O }  & $3.10\times10^{-18}$ & 0 & 9.31  & $4.90\times10^{-20}$ & $1.50\times10^{-19}$ & $2.86\times10^{-19}$  \\
\hline
{DMS + O $\rightarrow$ CH$_3$SO + CH$_3$ }  & $1.30\times10^{-17}$ & 0 & -3.40   & $5.91\times10^{-17}$ & $3.93\times10^{-17}$ & $3.10\times10^{-17}$ \\
{CH$_3$Cl + O $\rightarrow$ CH$_2$Cl + OH} & $1.74\times10^{-17}$ & 0 & 28.68  & $1.77\times10^{-23}$ & $8.77\times10^{-22}$ & $8.26\times10^{-21}$ \\
{CH$_3$Br + O $\rightarrow$ CH$_2$Br + OH }  & $2.21\times10^{-17}$ & 0 & 30.76  & $1.77\times10^{-23}$ & $8.77\times10^{-22}$ & $8.26\times10^{-21}$ \\
{CH$_3$I + O $\rightarrow$ CH$_3$ + IO }  & $6.19\times10^{-18}$ & 0 & -2.84  & $2.19\times10^{-17}$ & $1.56\times10^{-17}$ & $1.28\times10^{-17}$  \\
\tableline
\end{tabular}
\end{center}
\caption{Reaction rates with {H}, {OH}, and {O} of select Type III
  biosignature gases. Second order reaction rates in units of m$^3$
  molecule$^{-1}$ s$^{-1}$ are computed from the formula $k(T) = A
  (T/298)^n \exp(-E/RT)$ where $T$ is the temperature in K and $R$ is
  the gas constant ($R=8.314472\times10^{-3}$ kJ~mole$^{-1}$). The
  reactions rate are compiled from the NIST Chemical Kinetics
  Database. }
\label{tab-reactionrates}
\end{table*}

\begin{figure}[ht]
\begin{center}
\includegraphics[scale=.43]{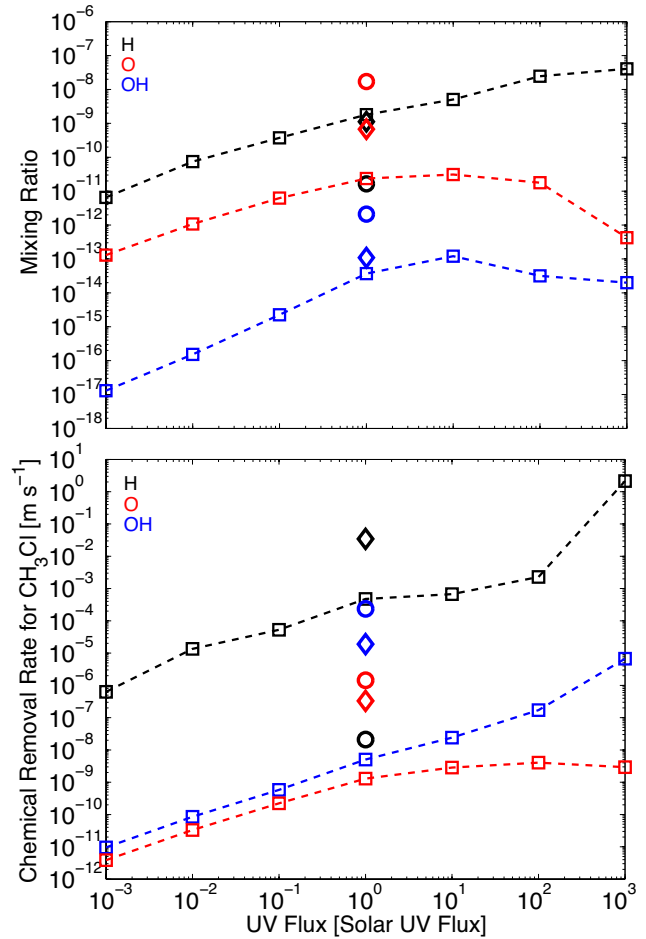}
\caption{Destructive power of reactive species (H, OH, and O) in a
  reduced atmosphere.  The atmosphere considered has 90\% H$_2$ and
  10\% N$_2$ by volume, with CO$_2$, CH$_4$, SO$_2$, and H$_2$S
  emission from its surface, for a 1~bar atmosphere on a planet with
  10~M$_\oplus$ and 1.75~R$_\oplus$.  Shown for comparison, are cases
  for an N$_2$-dominated atmosphere (diamond markers) and Earth's
  current atmosphere (circular markers).  {\it Top panel}: Mixing
  ratios of H, OH, and O as a function of UV flux. The mixing ratio of
  H exceeds the other reactive species OH and O. {\it Bottom panel}:
  The column-integrated chemical removal rates as a result of
  reactions with H, O, and OH, for which we have used CH$_3$Cl as an
  example. The removal rates are scaled by the steady-state mixing
  ratio of CH$_3$Cl to have a dimension of velocity.  This panel shows
  that removal by H is the dominant loss rate, and that the loss rates
  scale approximately linearly with UV flux incoming to the
  exoplanet atmosphere.  }
\label{fig:OHOHreduced}
\end{center}
\end{figure}

There is a very important point of comparison between the dominant
reactive species, H in H$_2$-rich atmospheres and OH in oxidized
atmospheres.  The concentrations of H and OH in the two different
types of atmospheres vary (see Figure~\ref{fig:OHOHreduced}
and Table~\ref{tab-reactionrates}, as well as more generally Table 4
in \citet{hu2012}).  This can be understood qualitatively because OH
is much more reactive than H. OH will react faster with any
atmospheric component than H, and so, for the same impinging stellar
UV flux, OH will build up to a lower atmospheric concentration than
H. The rate of removal of a biosignature gas by H or OH is a product
of the concentration and the reactivity. OH, with lower concentration
but greater reactivity will remove a biosignature gas at a similar
rate to H, which has a greater concentration but a lower
reactivity. In other words, while the mechanism of chemistry clearance
and the end products are different, the loss rate is fairly
similar. For more details of the formation and destruction of the
reactive species H, OH, and O in reduced and oxidized atmospheres, see
\citet{hu2012}.

\section{Results: Potential and Unlikely Biosignature Gases}
\label{sec-biosigresults}
We now turn to describe the potential and unlikely biosignature
gases in an H$_2$ atmosphere by their biosignature category.  The
biosignature categories developed in \citet{seag2013b} and summarized
in \S\ref{sec-biomassmodels} are an essential aide for calculations
because of the common formation pathways that belong to each biosignature
class.

We consider a planet with $10 M_{\oplus}$, 1.75 $R_{\oplus}$, an
atmosphere with 90\%~H$_2$ and 10\%~N$_2$ by volume.  The atmosphere
scenario is the hydrogen-rich case among the exoplanet benchmark
scenarios detailed in \citet{hu2012}, and we here outline the key
specifics. The planet surface pressure is 1 bar and the planet surface
temperature is 290~K. The temperature drops with increasing altitude
according to an adiabatic lapse rate, until reaching 160 K and is
prescribed as constant above. The semi-major axis of the planet’s
orbit is 1.4 AU if orbiting a sun-like star, 0.037~AU if orbiting an
M5V dwarf star, this is the consistent planet-star separation given
the atmospheric composition and the required surface temperature.  The
eddy diffusion coefficients are scaled up by a factor of 6.3 from
those measured in Earth’s atmosphere, in order to account for the
difference in the mean molecular mass. Important minor gases
considered are H$_2$O (evaporated from a liquid water ocean), CO$_2$
(about 100 ppm), and CH$_4$ and H$_2$S (emitted from
surface). Deposition velocities of H$_2$, CH$_4$ are assumed to be
zero, the deposition velocity of CO is $10^{-10}$~m~s$^{-1}$, and
deposition velocities of oxidants including O$_2$, O$_3$, and
H$_2$O$_2$ and sulfur species are assumed to be the values as on
Earth. See \citet{hu2012} for the rationale for these specifics and
for the description of carbon, oxygen, and sulfur chemistry in such an
H$_2$-dominated atmosphere.

The amount of UV flux on the planet from the star is critical to
  destroying biosignature gases and so we consider the same planet
  orbiting three different star types. The first star type is a
  sun-like star. The second star type is a weakly active 0.2~$R_\odot$
  M5V dwarf star, with EUV taken as that expected for GJ~1214b
  \citep{fran2013}. The third star type is a quiescent M star with no
  chromospheric and only photospheric radiation, again an
  0.2~$R_\odot$ M5V dwarf star (see
  \S\ref{sec-Mstarradiation}).  UV radiation received by the planet is
scaled according to the semi-major axis, and the stellar UV spectra
are from: the Air Mass Zero reference spectrum for the sun-like star
(http://rredc.nrel.gov/solar/spectra/am0/); the UV fluxes are 
from \citet{fran2013} for the weakly active M dwarf star (using the values
for GJ~1214b), and from simulated spectra of cool stars
\citep{alla1997} for the UV quiet M dwarf star (see Figure~\ref{fig:starflux}).

Whether or not a biosignature gas is detectable can be technique and
spectral feature dependent. The required atmospheric concentration
depends on the strength of a given absorption feature, and different
techniques are sensitive to different wavelength ranges. For example,
thermal emission detection sensitivity follows the planetary thermal
emission flux (approximately a black body peaking in the mid-IR),
whereas the transmission spectra sensitivity in the infrared follows
the thermal emission flux of the star (approximately a black body). An
illustrative example is NH$_3$ with a strong absorption feature at
10.3-10.7~$\mu$m suitable for planetary thermal emission but for
transmission a weaker absorption feature at 2.8-3.2~$\mu$m is more
easily detected than the 10~$\mu$m feature because of overall photon
fluxes of the star. For transmission spectra we avoid consideration of
sun-like stars because the observational signal (the planet
atmosphere-star area ratio) is too low \citep[e.g.,][]{kalt2009}.

Biosignature gas results are summarized for thermal emission
detectability (for sun-like and M dwarf stars) in
Table~\ref{tab-resultsTE} and for transmission spectra detectability
(for M dwarf stars only) in Table~\ref{tab-resultsTR}. Select promising
biosignature gases are shown via their thermal emission spectra for a
variety of atmospheres for intercomparison: CH$_3$Cl
(Figure~\ref{fig:CH3Clspectra}); DMS (Figure~\ref{fig:DMSspectra});
N$_2$O (Figure~\ref{fig:N2Ospectra}); NH$_3$ (see \citet{seag2013b}
Figure~2), and via their transmission spectra for H$_2$-dominated
atmospheres (Figure~\ref{fig:TransmissionSpectra}).

\begin{table*}[ht]
\begin{center}
\begin{tabular}{|l|l|l|ll|ll|ll|l|}
\hline 
Molecule & Mixing & Wave- & Surf. Flux & Biomass  & Surf. Flux     & Biomass & Surf. Flux & Biomass & Dominant \\ 
         & Ratio &  length          & Sun-Like     & Estimate & Active M  & Estimate & Quiet M  & Estimate & Removal \\ 
         & [ppm]         & [$\mu$m]   & [m$^{-2}$~s$^{-1}$]  & [g~m$^{-2}$] & [m$^{-2}$~s$^{-1}$]  & [g~m$^{-2}$] & [m$^{-2}$~s$^{-1}$]  & [g~m$^{-2}$] & Path \\
\hline 
Type I & & & & & & & & &\\ 
\hline 
NH$_3$ & 0.10 & 10.3-10.8 & 2.4$\times10^{15}$ & 4.0$\times 10^{-4}$ & 5.1$\times 10^{14}$ & 8.0$\times 10^{-6}$ & 8.2$\times 10^{5}$ & 9.5$\times 10^{-6}$ & photolysis\\ 
\hline 
Type III & & & & & & & & & \\ 
\hline 
CH$_3$Cl & 9.0 &13.0-14.2 & 1.0$\times 10^{17}$ & 2.8$\times 10^3$ & 2.9$\times 10^{15}$ & 77 & 4.7$\times 10^{11}$ & 0.013 & H \\ 
DMS & 0.10 & 2.2-2.8 & 4.2$\times 10^{19}$ & 190 & 1.8$\times 10^{19}$ & 82 & 2.4$\times 10^{13}$ & 1.1$\times 10^{-4}$& O \\ 
CS$_2$ & 0.59 & 6.3-6.9 & 8.7$\times 10^{17}$ & 5.5$\times 10^7$ & 3.6$\times 10^{17}$ & 2.3$\times 10^7$ & 5.9$\times 10^{11}$  & 37 & O \\ 
OCS & 0.10 & 4.7-5.1 & 2.5$\times 10^{15}$ & 1.3$\times 10^5$ &  1.0$\times 10^{14} $ & 5.5$\times 10^3$ & 1.3$\times 10^{10}$ & 0.67 & H \\
N$_2$O &  0.38  & 7.5-9.0   & 3.8$\times 10^{15}$ & --  & 5.4$\times 10^{14} $ & -- & 1.3$\times 10^{11}$ & -- & photolysis\\
\hline
\end{tabular}
\end{center}
\caption{Results for thermal emission spectra.  Potential biosignature
  gas required concentrations, related required biosignature gas
  surface fluxes (in units of molecules~m$^{-2}$~s$^{-1}$), estimated
  biomass surface densities, and the dominant removal path or destructive species.
  Results are given for three cases: for a planet orbiting a sun-like
  star, a weakly active M5V dwarf star (denoted``Active M'') and a
  quiescent M5V dwarf star (denoted ``Quiet M''). The planet
  considered has 10~M$_{\oplus}$, 1.75~R$_\oplus$, an atmosphere with
  90\%~H$_2$ and 10\%~N$_2$ by volume, with a surface temperature of
  290~K and a surface pressure of 1 bar.  Note that compounds with the
  removal path dominated by O, the required surface flux sensitively
  depends on the CO$_2$ emission/deposition.
}
\label{tab-resultsTE}
\end{table*}

\begin{table*}[ht]
\begin{center}
\begin{tabular}{|l|l|l|ll|ll|l|}
\hline
Molecule & Mixing & Wave- & Surf. Flux & Biomass & Surf. Flux & Biomass & Dominant       \\
         & Ratio &  length& Active M & Estimate & Quiet M & Estimate & Removal  \\
         & [ppm] & [$\mu$m]  & [m$^{-2}$~s$^{-1}$] & [g~m$^{-2}$] & [m$^{-2}$~s$^{-1}$] & [g~m$^{-2}$] & Path   \\  
\hline
Type I & & & & & & &\\
\hline
NH$_3$   & 11       & 2.8-3.2   & 5.5$\times 10^{16}$ & 1.1 & 8.8$\times 10^7$ & 1.8$\times 10^{-9}$& photolysis\\
\hline
Type III & & & & & & &\\
\hline
CH$_3$Cl &  10    & 3.2-3.4  & 3.2$\times 10^{15}$  & 8.6$\times 10^2$ & 5.2$\times 10^{11}$ & 1.4$\times 10^{-2}$ & H  \\
DMS      &  0.32  & 3.1-3.6  & 5.8$\times 10^{19}$  & 2.6$\times 10^2$ & 7.8$\times 10^{13}$ & 3.6$\times 10^{-4}$ & O   \\
CS$_2$   &  0.38   & 6.4-6.9  & 2.3$\times 10^{17}$  & 1.5$\times 10^7$ & 3.8$\times 10^{11}$ & 24 & O        \\
OCS      &  1.8   & 4.7-5.1  & 1.9$\times 10^{15} $ & 9.9$\times 10^4$ & 2.3$\times 10^{11}$ & 12 & H        \\
N$_2$O   &  11    & 3.8-4.1  & 4.8$\times 10^{15} $ & -- & 3.7$\times 10^{12}$  & -- & photolysis\\
\hline
\end{tabular}
\end{center}
\caption{Results for transmission spectra.  Potential biosignature gas
  required concentrations, related required biosignature gas surface
  fluxes (in units of molecules~m$^{-2}$~s$^{-1}$), estimated biomass
  surface densities, and the dominant removal path or destructive
  species.  Results are given both for two cases, a planet orbiting a
  weakly active M5V dwarf star (denoted ``Active M'') and a quiescent
  M5V dwarf star (denoted ``Quiet M''). The planet considered has
  10~M$_{\oplus}$, 1.75~R$_\oplus$, an atmosphere with 90\%~H$_2$ and
  10\%~N$_2$ by volume, with a surface temperature of 290~K and a
  surface pressure of 1 bar.  Note that a planet orbiting a sun-like
  star is not considered for transmission spectra because the overall
  detection signal is too low because of the small planet atmosphere annulus
  area to sun-like star area. Note that compounds with the removal
  path dominated by O, the required surface flux sensitively depends
  on the CO$_2$ emission/deposition. }
\label{tab-resultsTR}
\end{table*}

\subsection{Type I Biosignature Gases: Fully Reduced Forms}

We start by focusing on the Type I biosignature gases, gases generated
by reactions that extract energy from external, environmental redox
gradients. The most likely Type I metabolic product in an H$_2$-rich
atmosphere would be those in which non-hydrogen elements are in their
most hydrogenated form\footnote{Life might produce molecules with
  elements in intermediate redox states as life does on Earth.  In an
  H$_2$ atmosphere such molecules are likely to be photochemically
  hydrogenated.}. In a reducing environment, life captures
  chemical energy by reducing environmental chemicals. In the presence
  of excess hydrogen, the most energy life could extract from chemical
  potential energy gradients would be from converting elements from
  relatively oxidized compounds to their fully reduced form. An
  additional reason for focusing on Type I biosignature gas products
  that are in their most hydrogenated from is that they are likely
  long lived in an H$_2$-dominated atmosphere, because molecules in
  their most hydrogenated form cannot undergo any further reactions
  with H.

\subsubsection{Type I Biosignature Gas Overview}
The most reduced form of the most abundant non-metal elements, C, N,
O, P, S, H, Si, F, and Cl are CH$_4$, NH$_3$, H$_2$O, PH$_3$, H$_2$S,
H$_2$, SiH$_4$, HCl, and HF.  The most promising Type I biosignature
gas is NH$_3$, which is further described below
(\S\ref{sec-NH3result}). The other reduced molecules are unlikely
  biosignature gases for a variety of reasons.  Some (PH$_3$,
SiH$_4$), require energy input to make the reduced product from
geologically available materials, and so would not be produced by Type
I biosignature gas reactions.  Some are always present in their most
reduced form and so life cannot reduce them further (F, Cl). H$_2$S
and CH$_4$ are not viable for reasons discussed below, largely because
geological and biologically sources cannot be discriminated between.
In the case of H$_2$ and H$_2$O, they are naturally present in an
H$_2$-dominated atmosphere at relevant potentially-habitable planet
temperatures.

An aside about PH$_3$. We note that trace amounts of phosphine is
produced by some anaerobic ecologies on Earth \citep{glin2005}. It is
controversial whether the microorganisms in these environments are
making PH$_3$, or whether the bacteria are making acid which is
attacking environmental iron that contains traces of phosphide, and
this attack is making the phosphine gas \citep{roel2001}. Phosphine is
a potential biosignature in other highly reduced
environments. Phosphine is reactive and thermodynamically disfavored
over elemental phosphorus and hydrogen at Earth surface pressure and
temperature. Phosphine might be a Type I biosignature gas under
conditions of very high H$_2$ pressure, which would favour production
of PH$_3$ over elemental phosphorus.  Phosphine could be also be
produced as a Type III biosignature gas, analogous to reactive
signaling molecules such as NO or C$_2$H$_4$ on Earth.

\subsubsection{NH$_3$ as the Strongest Candidate Biosignature Gas in an H$_2$
Atmosphere}
\label{sec-NH3result}
NH$_3$ is the strongest candidate biosignature gas in a thin, H$_2$
atmosphere because, like O$_2$ in Earth's atmosphere, there is no
plausible geological or photochemical mechanism for producing high
concentrations on rocky planets with thin
  atmospheres (but c.f. the false positive discussion below). NH$_3$
is readily photolyzed in the upper atmosphere to yield N$_2$ and in
volcanic gases is thermally broken down at high temperatures.  The
triple bond of N$_2$ makes it extremely kinetically stable and so any
N in the atmosphere ends up being trapped as N$_2$.

We have therefore proposed NH$_3$ as a biosignature gas in an
H$_2$-rich atmosphere \citep{seag2013b}.  NH$_3$ is a good
biosignature gas candidate for any thin H$_2$-rich exoplanet
atmosphere because of its short lifetime and lack of geological
production sources. NH$_3$ as a biosignature gas is a new idea, and
one that is specific to a non-Earth-like planet. On Earth, NH$_3$ is
not a useful biosignature gas because, as a highly valuable molecule
for life that is produced in only small quantities, it is rapidly
depleted by life and is unable to accumulate in the atmosphere.
NH$_3$ is also a very poor biosignature gas on Earth because it is very 
soluble, so the trace amounts produced will stay dissolved in water and
not escape to the atmosphere.

The summary of the biosignature gas idea is that NH$_3$ would be
produced from hydrogen and nitrogen, in an atmosphere rich in both,
\begin{equation}
{\rm 3H_2 + N_2 \rightarrow 2NH_3.}
\end{equation}
This is an exothermic reaction which could be used to capture
energy. The industrial version of this reaction 
is called the Haber process for ammonia production at high temperatures;
hence we call such a planet a cold Haber
world.  We proposed that in an H$_2$-rich atmosphere, life could find
a way to catalyze the breaking of the N$_2$ triple bond and the H$_2$
bond to produce NH$_3$, and capture the energy released. 
In contrast, life on Earth solely fixes nitrogen in an energy-requiring
process. Energy capture would yield an excess of NH$_3$ over that needed by
life to build biomass, and so the excess would accumulate in the
atmosphere. Is a cold Haber World possible? We believe yes, based on
synthetic chemistry on Earth that can catalyze the breakage of each of
H$_2$ \citep{nish1998} and N$_2$ bonds \citep{yand2003, schr2011} at
Earth's surface pressure and temperature; what is not yet known is a
catalytic system that can break both at once.

We showed in \citet{seag2013b} that for an Earth-size, Earth-mass
planet with a 1 bar atmosphere of 75\% by volume N$_2$ and 25\% by
volume H$_2$ (including carbon species via a CO$_2$ emission flux), a
potentially detectable NH$_3$ atmosphere concentration of 0.1~ppm is
sustainable by a very reasonable biomass surface density of $9 \times
10^{-2}$ g~m$^{-2}$. This modest surface density corresponds to a
layer less than one bacterial cell thick. For comparison, the
phytoplankton that are the major contributor to Earth's oxygen
atmosphere are present in Earth's oceans at around 10 g~m$^{-2}$. For
interest, we note that standard printer paper is between 80 and 100
g~m$^{-2}$.

For an H$_2$-dominated atmosphere with 90\% H$_2$ and 10\% N$_2$ on a
planet with 10~M$_{\oplus}$, 1.75~R$_\oplus$ orbiting a sun-like star,
but all other parameters the same as the above, the viability of
NH$_3$ as a biosignature gas in a thermal emission spectrum still
holds based on a physically reasonable biomass surface density. We now
describe the estimate for the biomass surface density, using the
  Type I biomass equation (equation~(\ref{eq:TypeIBiomassModel})). We
  use the NH$_3$ source flux of $2.4 \times
  10^{15}$~molecule~m$^{-2}$~s$^{-1}$ (see Table~\ref{tab-resultsTE}).
  To compute $\Delta G$ we used $T=290$~K, and reactant and product
  concentrations at the surface in terms of partial pressures of N$_2$
  = 0.1, H$_2$ = 0.9, NH$_3 = 1.4 \times 10^{-7}$, giving $\Delta G =
  85.6$~kJ~mole$^{-1}$.
$P_{m_e} = 7.0 \times 10^{-6}$ kJ~g$^{-1}$~s$^{-1}$, we find a
biomass surface density of $4.9 \times 10^{-2}$~g~m$^{-2}$.  
Based on a reasonable biomass surface density, we therefore
  consider the NH$_3$ production flux to be viable in our Haber World
scenario. The global annual biogenic NH$_3$ surface emission in the
Haber World would be about $1100$~Tg~yr$^{-1}$. This is much higher
than the Earth's natural NH$_3$ emission at 10 Tg~yr$^{-1}$
\citep{sein2006}.  Comparing NH$_3$ production on the Haber world and
on Earth, however, is not valid. We are postulating that production of
NH$_3$ on the Haber world is a major source of metabolic energy for
life. A better emission rate comparison is to the biosignature gas
O$_2$ from Earth's principle energy metabolism,
photosynthesis. Earth's global oxygen flux is 200 times larger than
the Haber World's NH$_3$ surface emission, at about $2 \times 10^5$
Tg~yr$^{-1}$ \citep{frie2009}.

Turning to a weakly active M5V dwarf star, for the same fiducial planet,
the NH$_3$ surface flux required to sustain a detectable level of
atmospheric NH$_3$ in a thermal emission spectrum is $5.1 \times
10^{14}$~molecule~m$^{-2}$~s$^{-1}$ (see Table~\ref{tab-resultsTE}).
This value is about 5 times lower than the sun-like star example above
and therefore converts into a biomass estimate of about 5 times
smaller than the sun-like star example above, or about $1 \times
10^{-2}$~g~m$^{-2}$, due to the linear scaling of the problem.  For a
transmission spectra measurement for the same planet orbiting the
  same weakly active M dwarf star, the optimal wavelength range for
  detection is 2.8--3.2~$\mu$m, the required concentration is 11~ppm,
  and the required surface source flux is $5.5 \times
  10^{16}$~molecule~m$^{-2}$~s$^{-1}$, resulting in a surface biomass
  of 1~g~m$^2$. For this particular example, NH$_3$ in transmission
  vs. thermal emission, it is more difficult to detect NH$_3$ and
  hence a higher biomass is actually required for the same
  hypothetical type of life.

We emphasize that the NH$_3$ biosignature gas concept is not changed
for a planet with a massive (yet still ``thin'')
atmosphere with high surface pressure. As long as
the surface conditions are suitable for liquid water, NH$_3$ will not
be created by uncatalyzed chemical reactions.

 NH$_3$ is not immune to false postives. Although a rocky planet
  with a thin H$_2$-dominated atmosphere is unlikely to have an NH$_3$
  false positive, the challenge is in identifying the planetary (and stellar)
  characteristics. We describe three scenarios that could lead to the
nonbiological production of NH$_3$.

A rocky world with a hot surface of $\sim$820~K could generated NH$_3$
by the conventional Haber process if there is surface iron. Such a hot surface
temperature could presumably be ruled out from other observations.

A second scenario where NH$_3$ is naturally occuring is in the
atmospheres of gas giant planets or the so-called mini Neptunes. The
deep atmosphere may reach conditions where NH$_3$ can be formed
kinetically at the extremely high pressures necessary for NH$_3$
formation to be possible thermodynamically. On Jupiter, for example,
the $ {\rm H_2 + N_2 \rightarrow NH_3}$ reaction becomes significant
in comparison with vertical transport at about 1500~K, 1400~bar
\citep{prin1981}.  The only way we can discriminate between planets
with a massive envelope and a rocky planet with a thin atmosphere
where the pressures for 
the thermodynamic formation of NH$_3$ are not reached, is
with high-resolution spectra to assess the surface pressure
\citep{benn2012, benn2013}.

A third scenario for an NH$_3$ false positive is for planets with
outgassed NH$_3$ during evolution. The importance of ammonia for the
atmospheric evolution of Titan relates to primordial ammonia which
accreted with the ices of the moon and has not subsequently been
broken down either by internal heat (likely on a rocky planet) or by
external UV photolysis (which will rapidly break down any NH3 in the
atmosphere) \citep{shin2012}. In this case, ammonia is therefore
present as ice in the interior. This would be a challenging case to
ascertain, and illustrates of how an assignment of any gas as a
biosignature gas candidate has to be given a detailed probabilistic
assessment based on what we know about the planet concerned.

For any case, a quiescent M star with no chromospheric UV
emission---hence a planet with little to no destructive UV
flux---NH$_3$ can easily accumulate in the planet atmosphere and act
as a signficant false positive. NH$_3$ is destroyed by photolysis and
is very sensitive to the amount of UV radiation.

\subsubsection{CH$_4$ and H$_2$S as Unlikely Biosignature Gases}
CH$_4$ has been described at length as a possible biosignature gas on
early Earth and on exoplanets in oxidized atmospheres
\citep[e.g.,][]{hitc1967,desm2002}. This is despite the risk of a
geologically derived false positive, because it is believed that in an
oxidized environment geological production of CH$_4$ will be small,
and so if enough CH$_4$ is produced it may be attributed to life. This
is the case on Earth where at least 99\% of the atmospheric CH$_4$
derives directly from life or from industrial destruction of fossil
hydrocarbons formed from past life \citep{wang2004}.  However, the
1775~ppb concentration of CH$_4$ in Earth's atmosphere
\citep{solo2007} is not enough to be detected remotely with envisioned
space telescope capabilities.

CH$_4$ is a poor biosignature gas in an H$_2$-rich atmosphere because
it is both produced volcancially and is an end product of CO$_2$ 
photochemistry in
the atmosphere.  Terrestrial volcanic emission
rates of CH$_4$ and CO$_2$ would lead to substantial build-up of
CH$_4$ in H$_2$ dominated atmospheres. Even small amounts of outgassed
CO$_2$ will lead to an accumulation of CH$_4$ in the atmosphere,
because CH$_4$ has a very long lifetime in an H$_2$-rich atmosphere and
CH$_4$ would be produced by
\begin{equation}
{\rm 4H_2 + CO_2 \rightarrow CH_4 + 2H_2O}.
\end{equation}
More specifically, considering Earth's volcanic emission rates of
CH$_4$ and CO$_2$, and with deposition velocities of
10$^{-6}$~m~s$^{-1}$ for CO$_2$ and zero~m$^{-2}$~s$^{-1}$ for CH$_4$,
CH$_4$ will accumulate up to $\sim 10$ ppm in a 1 bar 90\% H$_2$, 10\%
N$_2$ atmosphere with a temperature profile similar to Earth's (Hu et
al. 2012).  Even in the case of no surface CH$_4$ emission, CO$_2$
emission into the same atmosphere would lead to the atmospheric
production and accumulation of CH$_4$ up to 5 ppm. This example is
intended to show that the false positive risk of CH$_4$ is so high in
a H$_2$-dominated atmosphere as to make CH$_4$ an implausible
biosignature gas.

H$_2$S is even more unfavorable than CH$_4$ as a biosignature gas in
an H$_2$-rich atmosphere because of the same geological false positive
issues as with carbon gases.  An added problem is the generation of
aerosols which may blanket any spectral features and the fact that the
H$_2$S spectral features are heavily contaminated by atmospheric water
vapor making them potentially difficult to detect \citep{hu2013}.

\subsection{Type II Biosignature Gases: No Viable Biosignature Gases}
\label{sec-TypeIIresults}
Type II biosignature gases are those produced by metabolic reactions
for biomass building. Biomass building on Earth primarily occurs by
photosynthesis, which has the dual goal of harvesting light energy to
use for metabolism and also for capturing carbon for biomass
building.

We have not identified any useful biosignature gases of Type II
in an H$_2$-rich, 1~bar atmosphere. Photosynthesis in a reduced environment
such as an H$_2$-dominated atmosphere would generate reduced byproduct
gases, which are not useful as biosignature gases because those
species are already expected to be present in their most reduced forms
in the H$_2$-dominated atmosphere.

The concept of photosynthesis on a planet with an H$_2$-dominated
atmosphere is nonetheless worth some discussion\footnote{see AbSciCon
  2008 abstract by N. Sleep
  http://online.liebertpub.com/doi/pdf/10.1089/ast.2008.1246, and \citet{pier2011} for a
  discussion of photosynthetic active radiation that reaches the surface
  under thick H$_2$ atmospheres.}, starting with a brief review of
photosynthesis in the familiar Earth environment.  Photosynthesis must
convert carbon from its environmental form, which is the form most
thermodynamically stable at surface temperatures and pressures, into
biomass. Biomass is of intermediate redox state \citep{bain2012}. The
key point therefore is that in an oxidized environment like Earth,
photosynthesis must reduce oxidized carbon (CO$_2$) and will generate
an oxidized byproduct.  On Earth environmental carbon is
captured in photosynthesis, producing O$_2$ as a byproduct,
\begin{equation}
{\rm H_2O + CO_2 \rightarrow CH_2O + O_2}.
\end{equation}
Here CH$_2$O represents biomass.

Photosynthesis, by definition, will have the same goal in an
H$_2$-rich environment as in an oxidized environment; to harvest light
energy and to build carbon-based biomass.  Because CH$_4$ is the most
thermodynamically stable gaseous form of carbon in this environment, photosynthesis would
oxidize the carbon in CH$_4$ and produce a reduced byproduct. The
lowest energy route is to directly split the CH$_4$, as,
\begin{equation}
\label{eq:photosyn1}
{\rm CH_4 + H_2O + X \rightarrow CH_2O + XH},
\end{equation}
where X is an atom that is oxidized in the environment, and has been
reduced to XH, consuming energy in the process, and CH$_2$O again represents
biomass. (We note that the oxidation state of the oxygen is not changed
in this process, unlike oxygenic photosynthesis on Earth, so formally
this is not splitting water even though water is involved.) 

The null result for biosignature photosynthetic biosignatures on an
H$_2$-dominated atmosphere is based on the point that most non-metals
(C, O, S, the halogens) are likely to be in their most reduced state
already on the surface of this world, and so cannot play the role of X
in the above described photosynthesis process.

One exception might have been hydrogen, which is oxidized in water and
methane, and so a possible photosynthetic reaction is
\begin{equation} 
\label{eq:photosyn2}
{\rm CH_4 + H_2O \rightarrow CH_2O + 2H_2},
\end{equation}
but again H$_2$ is not a useful biosignature gas
because it is already present in the H$_2$-dominated atmosphere. 

For completeness, we describe some other unlikely but interesting
possibilites for X and XH. Silicon, phosphorus and boron are likely to
be present as the oxidized minerals silicates, phosphates and borates
respectively, but using these as a sink for the electrons in
photosynthesis, for example in the reaction with silica to generate
silane,
\begin{equation} 
{\rm CH_4 + \frac{1}{2}  SiO_2 \rightarrow CH_2O  + \frac{1}{2} SiH_4},
\end{equation}
requires more energy than the reaction 
in equation~(\ref{eq:photosyn2}) under a range of conditions,
and so would represent a very inefficient way of generating
biomass. 

Reduction of a metal with a positive electrochemical potential
would be more energetically efficient, as for example in the reduction
of copper oxide to copper,
\begin{equation} 
{\rm CH_4 + 2Cu(O) \rightarrow CH_2O + 2Cu + H_2O},
\end{equation}
but produces no volatile product, and is dependent on a supply of
oxidized metal. (There are clear parallels with anoxygenic
photosynthesis on Earth for this type of reaction.) By contrast, the
reaction in equation~(\ref{eq:photosyn2}) is limited only by the supply
of methane, as life in water is not limited by the chemical
availability of water. In summary, photosynthesis in the reducing
environment will either generate H$_2$, which will not be
detectable in a hydrogen-dominated atmosphere, or will produce
non-volatile products, i.e. not products not in gas form
which by definition will not be detectable as atmospheric gases.

\subsection{Type III Biosignature Gases are Most Viable in Low-UV Environments}

Life produces many molecules for reasons that are not related to the
generation of energy, which we refer to as Type III biosignature
gases. The gases are produced for reasons such as
stress, signaling, and other physiological functions, and some of
these have already been discussed quantitatively in detail as
biosignature gases in oxidized atmospheres, (e.g., CH$_3$Cl \citep{segu2005}
and DMS and other sulfur compounds \citep{doma2011}).

The fate of Type III biosignature gas molecules depends on the level
of relevant reactive species in the atmosphere, and hence on stellar UV
flux. In low UV environments, some Type III gases can accumulate to
detectable levels. In the relatively high UV environments of sun-like
stars, in an H$_2$-rich atmosphere,  many Type III gases 
  could be rapidly driven to their most hydrogenated form, and 
  in some cases will not accumulate to detectable levels unless we
assume unrealistic production rates. In these extreme cases in a
  high UV environment, we would only be able to infer the presence of the
biosignature gas by detecting the end-product of photochemical attack,
which we call a bioindicator. Only in a few cases might bioindicators
be useful, because many are not spectroscopically active (and hence
not detectable) and others are indistinguishable from geological cases
as well (e.g., DMS will end up as CH$_4$ and H$_2$S, and N$_2$O will
end up as N$_2$ and H$_2$O.)

 We now show that Type III biosignature gas survival and hence
  plausibility depends highly on the UV flux level of the host
  star. We consider the three fiducial star types that differ in UV
  radiation levels: the sun-like star; the weakly active M5V dwarf
  star; and the quiescent M5V dwarf star (Figure~\ref{fig:starflux}).
  We consider the same model planet as above, a 10~M$_{\oplus}$,
  1.75~R$_\oplus$ planet with an atmosphere with 90\%~H$_2$ and
  10\%~N$_2$ by volume, with a surface temperature of 290~K and a
  surface pressure of 1 bar.  Results for the cases we modeled are
  listed for thermal emission spectra in Table~\ref{tab-resultsTE} and
  for transmission spectra in Table~\ref{tab-resultsTR}.  

Our first example of a Type III biosignature gas is methyl chloride
(CH$_3$Cl).  CH$_3$Cl is produced in trace amounts by many
microorganisms on Earth. The detectability of CH$_3$Cl in Earth-like
atmospheres in the low UV environment of UV quiet M stars has already
been studied by \citet{segu2005} and later as a potential biosignature
gas in more generalized oxidized atmospheres by \citet{seag2013b}.

Here for the first time we study CH$_3$Cl as a potential biosignature
gas in a thin H$_2$-rich atmosphere. For this, we go beyond previous work
not only by considering an H$_2$ atmosphere but also by using our
biomass estimate framework so as not to be constrained by terrestrial
bioflux production rates. We now show why CH$_3$Cl is a potential
biosignature gas in H$_2$ rich atmospheres in low UV
environments---because the amount of biomass to generate a detectable
concentration of CH$_3$Cl is physically plausible.  We use our biomass
estimate framework (\S\ref{sec-biomassmodels} and \citet{seag2013b}).

\begin{figure}[ht]
\begin{center}
\includegraphics[scale=.33]{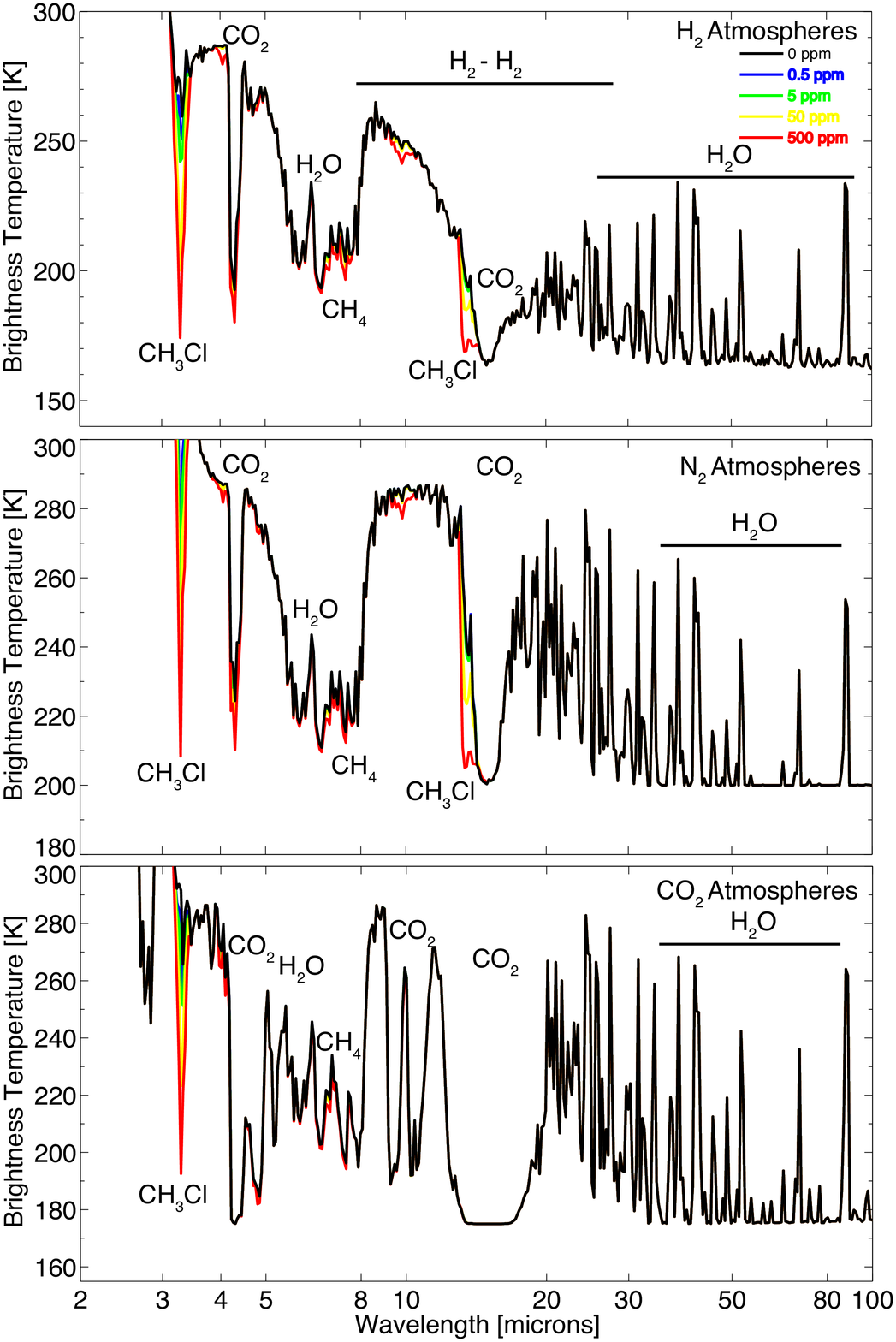}
\caption{Theoretical infrared thermal emission spectra of a super
  Earth exoplanet with various levels of atmospheric CH$_3$Cl in a
  1~bar atmosphere with a surface temperature of 290~K for a planet
  with 10~M$_{\oplus}$ and 1.75~R$_{\oplus}$.  From top to bottom, the
  panels show the spectra of H$_2$-, N$_2$-, and CO$_2$-dominated
  atmospheres, respectively, and the detailed compositions of these
  reference atmospheres are described in \S\ref{sec-biosigresults} for
  the H$_2$-dominated planet and in \citet{hu2012} for the N$_2$ and
  CO$_2$ dominated cases.  We find that over 5
  ppm of CH$_3$Cl is required for detection via thermal emission for
  H$_2$-, N$_2$-, and CO$_2$-dominated atmospheres.}
\label{fig:CH3Clspectra}
\end{center}
\end{figure}

Considering the thermal emission spectrum for our fiducial
planet with a 1 bar atmosphere of 90\%~H$_2$ and 10\%~N$_2$, a
spectral signature of 9 ppm is required for spectral detection using
our detection metric. This statement is for a spectral band feature in
absorption at 13.0-14.2~$\mu$m (see Figure~\ref{fig:CH3Clspectra});
this is the band accessible in an H$_2$ atmosphere, weaker than the
6.6-7.6~$\mu$m band that would be masked by H$_2$-H$_2$
collision-induced absorption.  In order to sustain an atmospheric
concentration of 9~ppm of CH$_3$Cl on our model planet in the
habitable zone for a sun-like, weakly active M5V dwarf star, and
  UV quiet M5V dwarf star, the surface bioflux production rate would
need to be $1.0 \times 10^{17}$~molecule~m$^{-2}$~s$^{-1}$ ($1.7
\times 10^{-7}$ mole~m$^{-2}$~s$^{-1}$), $2.9 \times
10^{15}$~molecule~m$^{-2}$~s$^{-1}$ ($4.8 \times 10^{-9}$
mole~m$^{-2}$~s$^{-1}$), $4.7 \times
10^{11}$~molecule~m$^{-2}$~s$^{-1}$ ($7.8 \times 10^{-13}$
mole~m$^{-2}$~s$^{-1}$), respectively.  Estimating the biomass with
equation~(\ref{eq:TypeIIIBiomassModel}) and with the lab rate at $6.17
\times 10^{-11}$~mole~g$^{-1}$~s$^{-1}$ \citep[see][]{seag2013b}, the
biomass surface density would need to be about 3000~g~m$^{-2}$,
80~g~m$^{-2}$, 0.001~g~m$^{-2}$ for each star-type respectively.  A
globally averaged density of 3000~g~m$^{-2}$ is likely too high, one
of 80~g~m$^{-2}$ is high but not impossible, according to terrestrial
biodensities \citep[see][]{seag2013b}.

The results show that CH$_3$Cl is a more viable biosignature gas in
low-UV as compared to high-UV environments. We emphasize that although
our estimates of biomass surface density for Type III biosignature
production are approximate, the resulting trend is robust.

For a spectral detection in transmission for our fiducial Earth
transiting an M5V star, the required concentration is about 10 ppm in
the wavelength range 3.2-3.4~$\mu$m. The surface bioflux and biomass
estimates for a weakly active and quiet star respectively are $3.2
\times 10^{15}$~molecule~m$^{-2}$~s$^{-1}$ and 900~g~m$^{-2}$ and $5.2
\times 10^{11}$~molecule~m$^{-2}$~s$^{-1}$ and 0.001~g~m$^{-2}$. The
required biomass surface density for the weakly active M star is
higher than the average surface biomass in Earth's oceans and the
biomass surface density for the quiet M5V dwarf star is much lower
than Earth's and very plausible, again emphasizing the trend that
low-UV radiation environments are more favorable for Type III
biosignature gas accumulation.

The different values for biosignature gas surface flux and for the
biomass estimates for transmission spectra as compared to thermal
emission spectra are in general due to either or both of longer
atmospheric pathlengths and different favorable wavelengths (depending
on the molecule of interest).

The fate of CH$_3$Cl in its destruction by H, is to end up in its
fully hydrogenated form, HCl, with the overall reaction as, 
\begin{equation}
{\rm CH_3Cl + H_2 \rightarrow CH_4 + HCl}.
\end{equation}
HCl could be a bioindicator. The HCl molecule is stable to further
photochemistry, because if it is photolyzed, the Cl atoms generated
will be predominately react with H to reform HCl.  HCl would not be
expected to be present in significant levels at atmospheric altitudes
for spectral detection without life taking non-volatile forms of Cl
and putting it into the atmosphere, because all geological sources are
non-volatile chlorides (such as NaCl) and any HCl that is volcanically
released would be efficiently rained out of the troposphere.  The
limiting problem is that the HCl spectral features are too weak to be
detectable and are likely to be contaminated by CH$_4$ in the 3 to 4
$\mu$m range.

\begin{figure}[ht]
\begin{center}
\includegraphics[scale=.33]{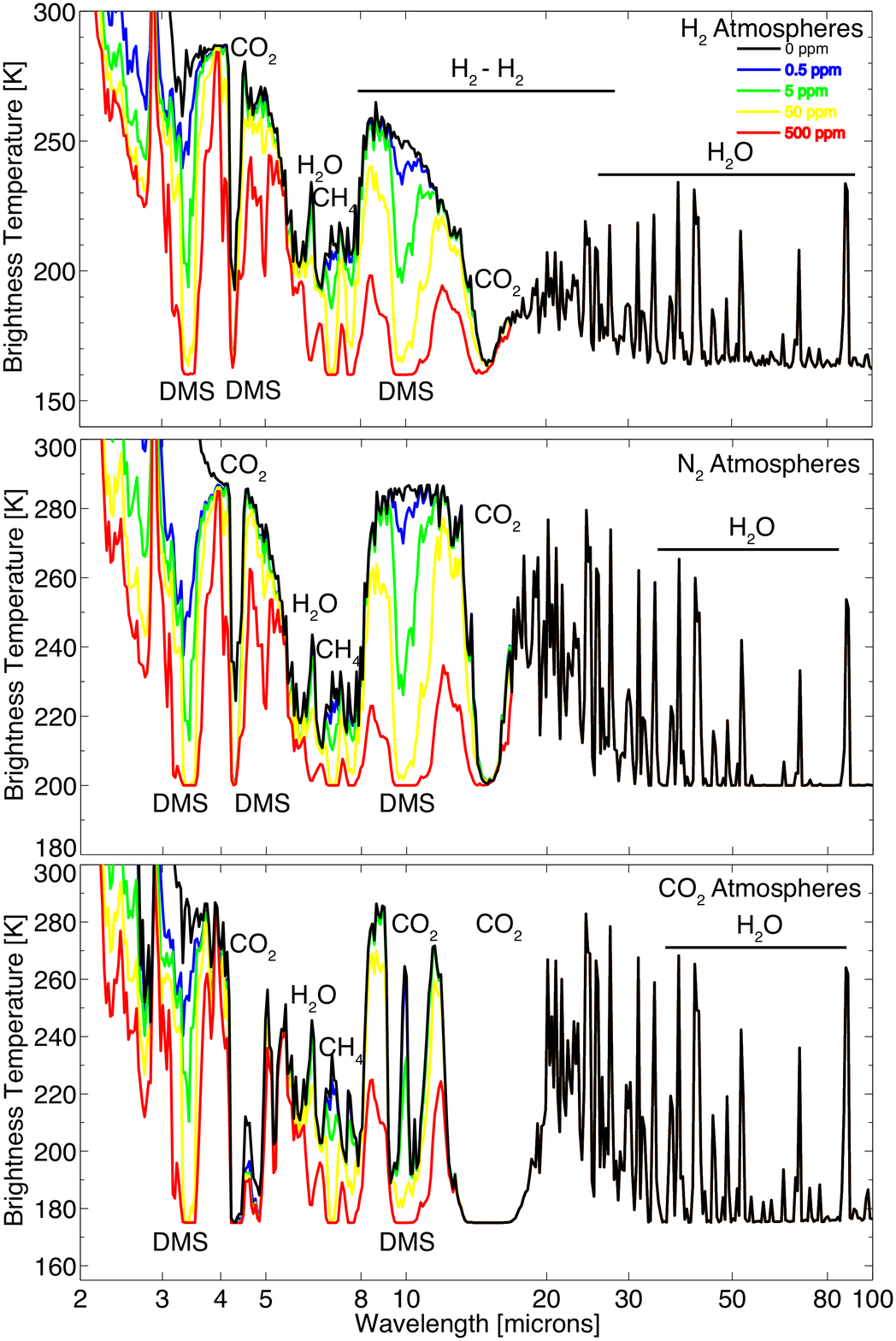}
\caption{Theoretical infrared thermal emission spectra of a super
  Earth exoplanet with various levels of atmospheric DMS in a 1~bar
  atmosphere with a surface temperature of 290~K for a planet with
  10~M$_{\oplus}$ and 1.75~R$_{\oplus}$. From top to bottom, the
  panels show the spectra of H$_2$-, N$_2$-, and CO$_2$-dominated
  atmospheres, respectively, and the detailed compositions of these
  reference atmospheres are described in \S\ref{sec-biosigresults} for
  the H$_2$-dominated planet and in \citet{hu2012} for the N$_2$
  and CO$_2$ dominated cases.  We find that 0.1 ppm of
  DMS is required for future detection via thermal emission for H$_2$, N$_2$,
  and CO$_2$-dominated atmospheres.}
\label{fig:DMSspectra}
\end{center}
\end{figure}

As a second Type III biosignature gas example we consider dimethyl
sulfide (DMS). DMS has been studied before in oxidizing
atmospheres by \citet{doma2011} who concluded that DMS itself is not a
potentially detectable biosignature gas in oxidized atmospheres under
sun-like UV radiation, but one of its photolytic breakdown products
ethane is detectable (we call this a ``bioindicator'' gas). Using the
same atmosphere and framework as the above CH$_3$Cl example, for
thermal emission spectra we find a mixing ratio required for detection
of 0.1 ppm in the 2.2-2.8~$\mu$m band (see
Figure~\ref{fig:DMSspectra}). Via photochemistry, this mixing ratio
corresponds to a surface flux in our fiducial H$_2$-dominated
atmosphere for a sun-like star, weakly active M5V dwarf star, and
quiet M5V dwarf star as $4.2 \times
10^{19}$~molecules~m$^{-2}$~s$^{-1}$ ($6.9 \times
10^{-5}$~moles~m$^{-2}$~s$^{-1}$), $1.8 \times
10^{19}$~molecules~m$^{-2}$~s$^{-1}$ ($3.0 \times
10^{-5}$~moles~m$^{-2}$~s$^{-1}$), $2.4 \times
10^{13}$~molecules~m$^{-2}$~s$^{-1}$ ($4.1 \times
10^{-11}$~moles~m$^{-2}$~s$^{-1}$), respectively.  Using a DMS lab
production rate of $3.64 \times 10^{-7}$~moles~g$^{-1}$~s$^{-1}$
\citep{seag2013b}, we come up with am implied biomass surface density
estimate for the three star types of about 200~g~m$^{-2}$,
100~g~m$^{-2}$, 10$^{-4}$~g~m$^{-2}$, respectively. The first two
values are high, but physically plausible as compared to Earth biomass
surface density ranges. For transmission spectra the numbers are about
a factor of two higher for the weakly active and quiet M5V dwarf star
(see Figure~\ref{fig:TransmissionSpectra} and
Table~\ref{tab-resultsTR}).

The DMS results show again that the lowest UV environment is most
favorable.  There are two other other relevant points related to
  DMS appearing to be a favorable biosignature gas in each of the
  three UV radiation environments studied. The first point is that
  gases destroyed by reaction with O (as opposed to gases destroyed by
  reactions with H) show a similar surface flux requirement between
  the sun-like and weakly active M dwarf star. This is because the
  release of O from CO$_2$ photolysis is largely driven by Lyman-alpha
  emission which are similar at the habitable zones for the sun-like
  and weakly active M dwarf star used in this study
  (Figure~\ref{fig:starflux}).

The second point is that the high $R_{\rm lab}$ values used for DMS,
and hence the low biomass surface density estimates, are a result of
the biology of DMS production. On Earth, DMS is the waste product of
consumption of DMSP by marine organisms consuming marine
plankton. DMSP is accumulated in large amounts by some marine
species. Thus organisms that generate DMS do not have to invest their
own resources to make DMS, and so are not limited to how much they can
make. Maximal production rates are therefore very high. This is
discussed further in \citet{seag2013b}.

In terms of a bioindicator, DMS will react with H$_2$ to generate
CH$_4$ and H$_2$S.  Neither is a useful bioindicator as CH$_4$ and
H$_2$S are expected to be present in the atmosphere naturally.  This
is in contrast to oxidized atmospheres, where ethane may be a
bioindicator gas, as the expected to be the end product of DMS
photodestruction by combination of methyl radicals generated from
attack of O on DMS \citep{doma2011}.

As a third and fourth Type III biosignature gas example, we used
CS$_2$ and OCS. For these two gases we find the same trend as the
other Type III biosignature gases, that is in a low UV environment the
biosignature gases can accumulate (see Tables~\ref{tab-resultsTE} and
\ref{tab-resultsTR}). The biomass estimates (as a plausibility check)
are too high for the sun-like and weakly active M dwarf star
environments to be plausible as compared to terrestrial biomass
surface density values. OCS in the UV environment of a weakly active
  M dwarf star may be an exception with an estimate at the upper limit
  of plausibility.

As a fifth example we describe N$_2$O. On Earth N$_2$O is a Type I
biosignature gas produced by nitrifying bacteria.  N$_2$O is not
likely to be produced in a thin H$_2$-rich atmosphere because there is
unlikely to be much nitrate available. Here we explain further. N$_2$O
has been suggested as a biosignature gas in Earth’s atmosphere
\citep{segu2005}. N$_2$O is a Type I biosignature gas formed by two
processes on Earth---the oxidation of ammonia by atmospheric oxygen
and the reduction of nitrate in anoxic environments,
\begin{eqnarray}
{\rm 2NH_3  + 2O_2 \rightarrow N_2O + 3H_2O}, \\
{\rm NO_3^- + H \rightarrow N_2O + H_2O}.
\end{eqnarray} 
Analogous reactions on a hydrogen-dominated world would be the
reduction of nitrate by atmospheric hydrogen
\begin{equation} 
{\rm NO_3^- + H_2 \rightarrow N_2O + H_2O + OH^-},
\end{equation}
or the oxidation of ammonia by a geologically derived oxidant.

Nitrate is formed on Earth by oxidation of NO generated by lightning
in Earth’s oxygen-rich atmosphere, or by biological
processes---neither are likely in an H$_2$-rich environment, so it is
not clear whether nitrate reduction is a useful energy source in a
world with an atmosphere rich in H$_2$. Ammonia oxidation requires a
strong oxidizing agent, which again is likely to be missing from the
environment. 

N$_2$O as a Type I biosignature gas, therefore, seems unlikely,
although not impossible, from very rare environments in which
there are oxidized nitrogen species generated geochemically.

\begin{figure}[ht]
\begin{center}
\includegraphics[scale=.33]{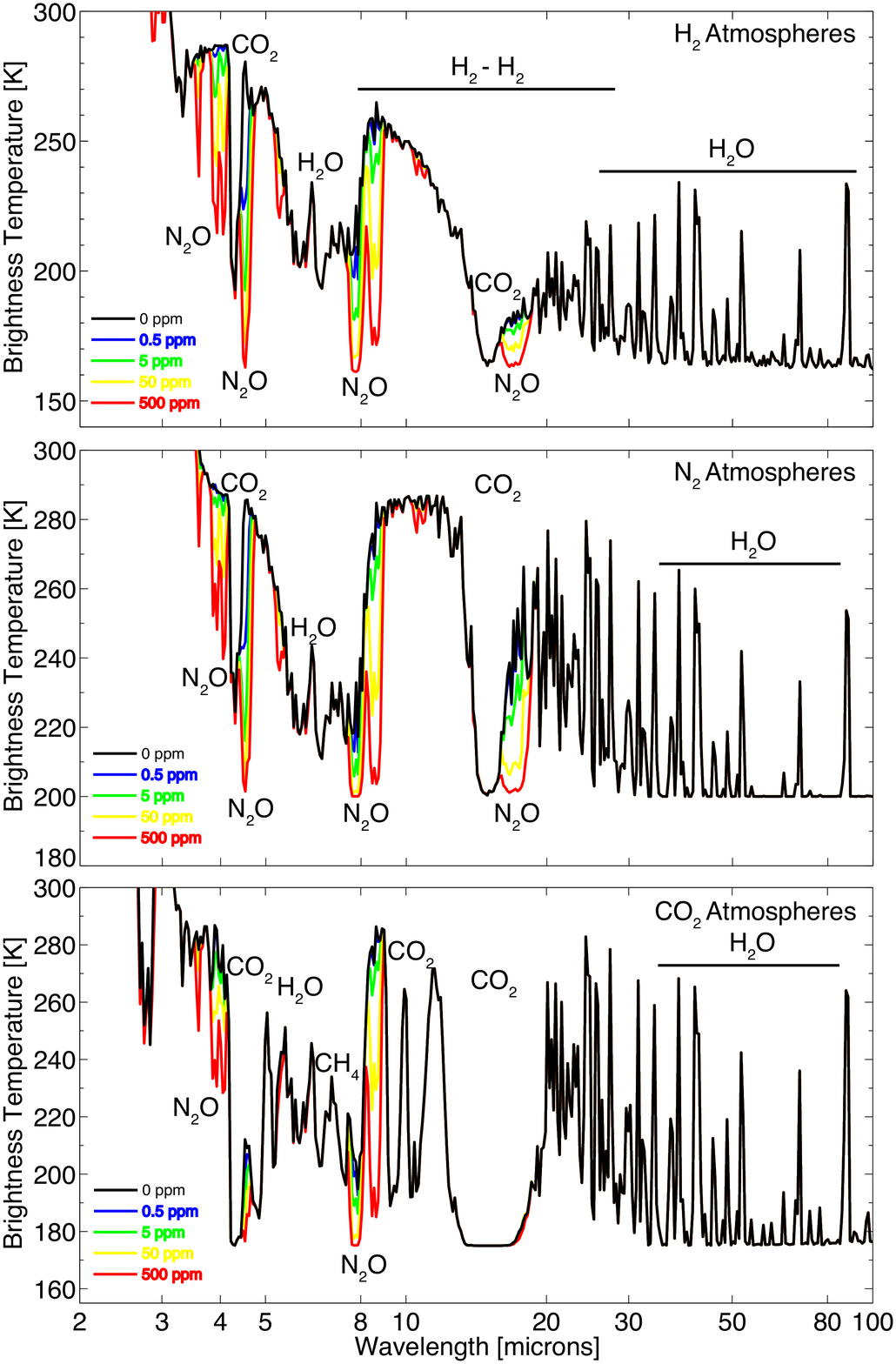}
\caption{Theoretical infrared thermal emission spectra of a super
  Earth exoplanet with various levels of atmospheric N$_2$O in a 1~bar
  atmosphere with a surface temperature of 290~K for a planet with
  10~M$_{\oplus}$ and 1.75~R$_{\oplus}$.  From top to bottom, the
  panels show the spectra of H$_2$-, N$_2$-, and CO$_2$-dominated
  atmospheres, respectively, and the detailed compositions of these
  reference atmospheres are described in \S\ref{sec-biosigresults} for
  the H$_2$-dominated planet and in \citet{hu2012} for the N$_2$ and
  CO$_2$ dominated cases.  We find that about 0.4 ppm of
  N$_2$O is required for future detection via thermal emission for H$_2$, N$_2$,
  and CO$_2$-dominated atmospheres.}
\label{fig:N2Ospectra}
\end{center}
\end{figure}

\begin{figure*}[ht]
\begin{center}
\includegraphics[scale=.4]{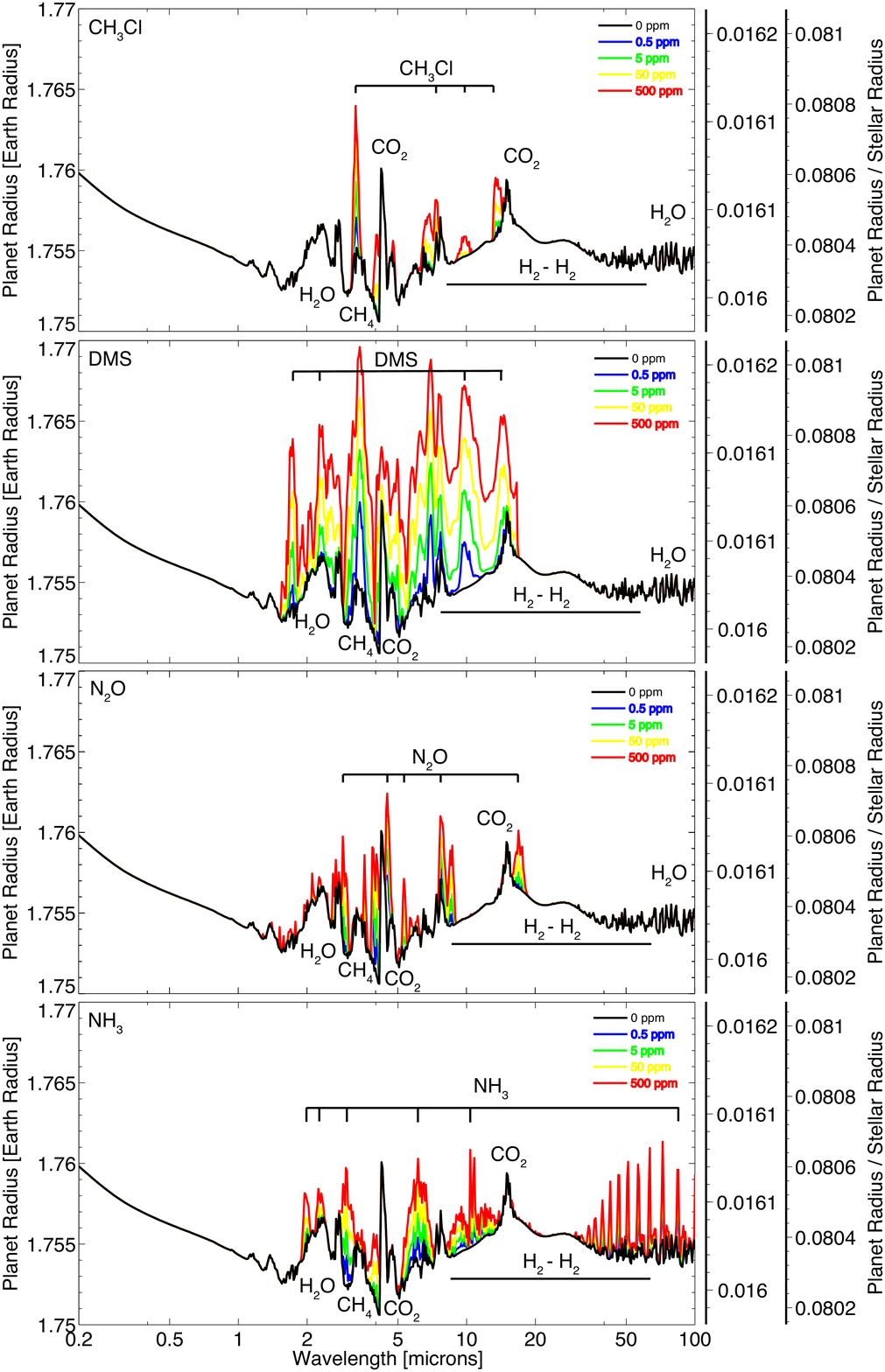}
\caption{Theoretical transmission spectra for potential biosignature
  gases in a 10 $M_\oplus$, 1.75 $R_\oplus$, 1 bar atmosphere composed
  of 90\% H$_2$ and 10\% N$_2$ and with a surface temperature of
  290~K.  Potential biosignature gases, including CH$_3$Cl, DMS,
  N$_2$O, and NH$_3$, have spectral features in infrared wavelengths
  from 1 to 10 $\mu$m, making these gases detectable at various
  atmospheric mixing ratios (see Table~\ref{tab-resultsTR}).}
\label{fig:TransmissionSpectra}
\end{center}
\end{figure*}

N$_2$O, however, could be a Type III biosignature gas as NO is for
some organisms on Earth. We have calculated the surface fluxes for a
detectable amount of N$_2$O in a thin, H$_2$-dominated atmosphere and find
relatively low required surface fluxes.  
The reason
is that N$_2$O is destroyed by photodissociation, a slower
rate than by reaction with H. N$_2$O may therfore be a
plausible biosignature gas candidate, even in an atmosphere subject to
strong UV radiation (see Figures~\ref{fig:N2Ospectra} and
\ref{fig:TransmissionSpectra} and Tables~\ref{tab-resultsTE}) and
\ref{tab-resultsTR}). A biomass estimate (as a plausibility check) is
not possible for N$_2$O, as it is only known as a Type I biosignature
on Earth (and so therefore Type III $R_{\rm lab}$ rates are not available).

\section{Discussion}
\label{sec-Discussion}

\subsection{What Constitutes an H$_2$-Dominated Atmosphere?}

We have calculated biosignature gas accumulation in an atmosphere with
90\% H$_2$ and 10\% N$_2$ by volume. A super Earth exoplanet
atmosphere can have many other gas species.  The concentration of the
major destructive species, H, O, and OH will depend on the amounts of
these other gas species.

As an example, we explore the changing effect of the reactive species
in an H$_2$-dominated atmosphere for different UV flux levels, based
on the surface flux levels of CO$_2$
(Figure~\ref{fig:mixingratiosvaryingCO2}). A few key points are as
follows.  The H abundance is almost not affected by the CO$_2$ mixing
ratios ranging from 10$^{-8}$ to 10$^{-2}$. The O abundance depends on
both CO$_2$ and UV levels, such that both a high CO$_2$ level and a
high UV flux lead to high atmospheric O. Only in extreme cases (e.g.,
H$_2$-dominated atmospheres with $>$1\% CO$_2$, shown by dashed
lines), the abundance of O may be very close to the abundance of H.
The OH abundance depends on a complex source-sink network, ultimately
driven by H$_2$O and CO$_2$ photolysis. Notably, the amount of H is
always at least 4 orders of magnitude higher than the amount of OH.

The effect of changing the H$_2$ mixing ratio and the addition and
variation of other active gases on the H concentration will need to be
considered on a case-by-case basis as they will react not only with H
and OH but also with other gas species.

\subsection{Can Super Earths Retain H$_2$-Dominated Atmospheres?}
\label{sec-Hescape}

Whether or not a super Earth planet can retain H$_2$ stably from
atmospheric escape is not known. Although many models and studies for
exoplanet atmospheric escape exist \citep[see e.g.,][and references
  therein]{lamm2012}, the permanent limitation is that there are too
many unknowns to provide a definitive and quantitative statement on
which planets will retain H$_2$.  One of the challenges is the unknown
history and present state of the host star's EUV flux. Another major
challenge is the defining the mechanism for atmospheric escape for a
given exoplanet, for example whether or not the regime of rapid
hydrodynamic escape was reached in a planet's history or which
non-thermal mechanism, if any, came into a dominant role \citep[see
  Table 4.1 and references therein in][]{seag2010}.  With an unknown
initial atmospheric reservoir and an unknown present atmospheric
composition, the regime and type of atmospheric escape is difficult to
impossible to identify.

Some super Earths will have been formed with atmospheres with H$_2$,
based on both theoretical and observational evidence.  Theoretically,
planetary building blocks containing water-rich minerals that can
release H \citep{elki2008,scha2010}. Observationally, a large number
and variety in radius of {\it Kepler} mini-Neptunes that must have H
or an H/He envelope to explain their radii. So either from outgassing
or nebular capture of gases, some super Earths should have started out
with H$_2$-rich atmospheres and those with high enough gravity and low
enough temperatures and/or with magnetic fields should be able to
retain the H$_2$ \citep[e.g.,][]{pier2011}. Observational detection of
H$_2$-rich atmospheres will ultimately be needed to confirm the
scenario of thin H$_2$-dominated atmospheres on super Earths.

\subsection{Upper Temperatures for Life}
\label{sec-lifeTlimits}
Super Earths with H$_2$-dominated atmospheres can have surface
temperatures hotter than Earth due to an H$_2$ greenhouse effect from
H$_2$-H$_2$ collision-induced opacities
\citep{bory2002,pier2011}. While the hypothetical planets we have
described in this paper were constructed to have 1 bar atmospheres
with Earth-type surface temperatures, many H$_2$-dominated planet
atmospheres are likely to have hotter surface temperatures than Earth,
even for planets orbiting beyond 1~AU of their host stars.

An important question for understanding the potential of biosignature
gases on a planet with an H$_2$-dominated atmospheres is therefore,
``how hot a planet can be and still sustain life?'' On Earth,
organisms that grow at 395~K are known \citep{lovl2003, taka2008} and
have been cultured in the lab at elevated pressures equal to in situ
pressures. Furthermore, proteins can function at 410~K to 420~K
\citep{tana2006, sawa2007, unsw2007} motivating a consensus that
life at 420~K is plausible \citep{demi1993, cowa2004}.

Life might exist at temperatures even higher than 420~K.  The main
argument for a maximum temperature for life involves the temperature
at which the basic building blocks of life (DNA, proteins,
carbohydrates, and lipids) break down.  Many of the component
chemicals of life, including DNA, many of the amino acids that
make up proteins, and many of the key metabolites that allow life’s
biochemistry to function are rapidly chemically broken down above
470~K \citep[e.g.,][]{cowa2004} The maximum temperature at which life
could exist therefore may lie between 420~K and 470K.

\subsection{What Surface Pressure is too High?}
\label{sec-masiveatm}

Many super Earth atmospheres will be much more massive than the 1~bar
atmosphere on Earth. For temperatures suitable for the existence of
liquid water (see \S\ref{sec-lifeTlimits}), the surface pressure could
be as high as 1000 bar or higher \citep{wagn2002}.  There are three
key points to show that the high surface pressures does not destroy
the biosignature gases before they can reach the high atmosphere.

``Can life generate potentially detectable biogsignature gases under a
massive atmosphere?''  The answer is yes, provided the surface
temperature is compatible with life, then in principle life can
survive and generate biosignature gases. The chemistry described in
this paper still holds under a massive atmosphere, because the
photochemical destruction occurs above 1~mbar.  Furthermore, we showed
in \citet{seag2012} that the biomass surface density estimates are
unchanged under a massive atmosphere as long as the photochemical loss
rate dominates. For biosignature gases whose loss is dominated by
deposition at the surface (i.e. are absorbed by the surface), then the
biosignature source flux and hence biomass surface density will scale
linearly with planetary atmosphere mass.

The second key question is, ``Can the high density and pressures on
the surface under a massive atmosphere generate false positives?''
The answer is almost entirely no, because we have shown that it is
largely the Type III biosignature gases that are viable biosignature
candidates.  Recall that Type III gases are those produced for reasons
other than energy extraction, and not from chemicals in the
environment. Therefore they are unlikely to be the product of
nonbiological chemistry; a statement that holds even under a massive
atmosphere. A comment related to NH$_3$ potential false positive is in
order.  For NH$_3$ to be generated from N$_2$ and H$_2$ kinetically,
the temperature has to be well above any temperature compatible with
life for any pressure where water is liquid, extrapolating from the
known fact that at 300~bar and 673~K N$_2$ and H$_2$ still need a
catalyst to be converted to NH$_3$, and higher pressures shouldn't
change this.  The false positive risk is instead in detecting NH$_3$
without being able to identify the surface as cold enough not to
possibly generate NH$_3$ kinetically.

The third key question is, ``Will the high surface pressures enable
fast chemical reactions that destroy the biosignature gases 
generated at the surface?'' The answer is no, for pressures under
about 1000 bar. In principle, if the upward diffusion or convective
motions bring the biosignature gas to higher altitudes faster than the
gas is destroyed by kinetic reactions, the higher surface pressures
will not interfere with biosignature gas accculumation in the
atmosphere.

Up to 1000 bar, reaction rates extrapolated from low-pressure kinetic
experiments should be valid to an order of magnitude. The additional
caveat is that low-pressure gas kinetics are usually measured at high
temperature and we are extrapolating to high pressure and low
temperature.  For example, at 1000 bar and 300~K, we estimate the half
life of hydrogenation of CH$_3$Cl to be 6.0$\times 10^{11}$~yr, the
half life of DMS to be $3 \times 10^{10}$~yr, and the half-life of
N$_2$O at $1 \times 10^{20}$~yr.  These numbers are based on
thermochemical equilibrium of H$_2$ and H based on the relative Gibbs
free energy of formation of atomic hydrogen and T and P
\citep[e.g.,][]{borg2009}.  The overview is that there is very little
free H at high pressures and low temperatures since in the absence of
UV, H atoms will be generated almost entirely thermochemically.

At pressures above 1000 bar we are less confident that chemistry can
be extrapolated even qualitatively from low pressure experiments.  By
1000~bar, most gases will have densities approaching those of their
liquids.  Increases in pressure will force molecules closer than their
van der Waals radii, directly altering molecular orbitals and reaction
pathways. Below $\sim$1000~bar we can consider molecules separate entities
and we can still consider the molecules’ chemistry to be qualitatively
similar to that of their dilute (ideal) gas state, and hence
order-of-magnitude extrapolations from low pressure gas chemistry is
justifiable. 

\subsection{Can we Identify Exoplanets with H$_2$-Dominated Atmospheres
that are Potentially Habitable?}

Given the argument that life can generate biosignature gases on a
planet with an H$_2$-rich atmosphere, but that the surface must have
the right temperatures, how can we identify suitable planets for
further study?  The problem is that super Earths are observed with a
wide range of masses and sizes, and we can anticipate that a range of
atmosphere masses will also exist.  A challenge is presented in
observational atmosphere studies because we can only ``see'' to an
optical depth of a few and this limiting optical depth can be
reached well above any surface for a thick atmosphere. In many cases
surface conditions cannot be probed. Ideally,
high-resolution spectra can be used to tell whether or not the
atmosphere is thick or thin (i.e. whether or not one can
observe down to the planetary surface), based on the shape of the spectral
features, as described in detail in \citet{benn2012,benn2013}.

We support the search for biosignature gases regardless of being able
to classify a planet as habitable, because identifying biosignature
gas molecules may be more easily attainable than high-spectral resolution
characterization of a super Earth atmospheric spectrum. 
That said, where possible, planetary radii can be used to discriminate
planets worthy of followup since those with small enough radii can be
identified to likely have thin atmospheres and those with radii large
enough to have massive H$_2$ or H$_2$/He envelopes are unsuitable
\citep{adam2008}.

\subsection{The M Star UV Radiation}
\label{sec-Mstarradiation}
 
Biosignature gases can more easily accumulate in a low-UV radiation
environment as compared to a high-UV radiation environment because the
UV creates the destructive atmospheric species. We have shown this for
H$_2$ atmospheres in this paper and \citet{segu2005} have shown this
for Earth-like planet atmospheres.

Whether or not truly UV quiet M dwarfs exist and if UV activity is 
correlated with photometric stability is unclear. Recently
\citet{fran2013} observed a small sample of six planet-hosting M dwarf
stars with HST observations at far-FUV and near-UV wavelengths and
found none to be UV quiet. Other studies with much larger numbers of M
dwarf stars are ongoing, including some with UV emission from GALEX
(A. West, 2013 private communication). A general understanding is that
magnetic activity, as traced by H$\alpha$ in M dwarfs decreases with
age but that M dwarfs appear to have finite activity lifetimes such
that the early-type M dwarfs (M0-M3) spin down quickly with an
activity lifetime of about about 1 to 2 Gyr whereas later-type M dwarf
stars (M5-M7) continue to spin rapidly for billions of years
\citep{west2006, west2008}.

For the time being UV (that is the relevant FUV and EUV) radiation emitted
by stars of interest is not usually measured or theoretically known and so we have worked with three different UV-radiation environments (Figure~\ref{fig:starflux}).

A relevant point for quiet M stars with extremely low UV radiation (if
they exist) is that false positives for biosignature gases destroyed
by photolysis may also more easily accumulate.  This is relevant for
NH$_3$, for which the lifetime in our fiducial H$_2$ planet
atmosphere for a planet orbiting a quiet M star is about 1.4 Gyr,
according to our photochemistry models. This means that the false
positive risk that comes from primordial NH$_3$ would be
high\footnote{NH$_3$ requires on average two photons for destruction
  hence its UV lifetime is particularly sensitive to UV
  levels.}. Related issues with other gases that are primarily
destroyed by photolysis (and not destruction by reactions with H and
O) should be investigated.

\subsection{Detection Prospects}

Is there any hope that the next space telescope, the {\it James Webb
Space Telescope} could be the first to provide evidence of biosignature
gases? Yes, if—--and only if—--every single factor is in our favor. 

First, we need to discover a pool of transiting planets orbiting
nearby (i.e., bright) M dwarf stars. Second, the planet atmosphere should
preferably have an atmosphere rich in molecular hydrogen to increase
the planetary atmosphere scale height.  Third, the M dwarf star needs to be
a UV quiet M dwarf star with little EUV radiation. Fourth, the planet must have
life that produces biosignature gases that are spectroscopically
active.

Several biosignature gases, if they exist, are detectable with tens of
hours of JWST time, based on our detection metric.  Although our
detection metric assumes photon noise is the limiting factor, many
more detailed simulations of JWST detectability show that spectral
features of similar magnitude are detectable
\citep[e.g.,][]{demi2009}.

For detecting molecules using transmission spectroscopy, the
background exoplanet atmosphere dominated by H$_2$ or CO$_2$ has
little effect on the detectability of the biosignature gases of
interest we studied. This is because the transmission observations are
better performed in the near infrared than in the mid infrared because
of a higher stellar photon flux at near-infrared wavelengths, and the
contamination effects of either of the dominant CO$_2$ absorption or
collision-induced H$_2$-H$_2$ absorption are minimal in the near
infrared. As long as all of the biosignature gases of interest have
features in the near infrared (see
Figure~\ref{fig:TransmissionSpectra} for the spectral features), these
gases may be detected for atmospheres with any level of CO$_2$. The
key issue here, instead of spectral contamination, is the mean
molecular mass. The depth of the transmission spectral feature is 1
order-of-magnitude larger for H$_2$-dominated atmospheres compared
with CO$_2$-dominated atmospheres (see the scale height in
equation~(\ref{eq:scaleheight})). 

For detecting molecules via thermal emission with future direct
imaging techniques, one may expect the CO$_2$ or H$_2$-H$_2$
contamination to be important because thermal emission of the planet
peaks in the mid infrared where CO$_2$ and H$_2$-H$_2$ contamination
is most substantial. For individual gases, however, there are often
multiple absorption bands to mitigate this issue.
Similarly, a variety of wavelength ranges are usually available to
choose from for the other biosignature gases of interest studied in
this paper (as shown in Figures~\ref{fig:CH3Clspectra} to
\ref{fig:TransmissionSpectra}).

At this point we conclude by emphasizing a related point that the
  plausibility of a specific biosignature gas depends on the planet
  surface gravity, atmospheric pressure, and other characteristics,
  because such characteristics affect which atmospheric wavelength
 ``windows'' are
  most favorable. Individual planets and their atmospheres should be
  considered on a case-by-case basis.

\section{Summary and Conclusion}
\label{sec-summary}
We have provided a ``proof of concept'' that biosignature gases can
accumulate in exoplanets with thin H$_2$-dominated atmospheres. We used a model
atmosphere including a detailed photochemistry code and also employed
a biomass model estimate to assess plausibility of individual
biosignature gases. We considered a fiducial super Earth of 10
M$_{\oplus}$ and 1.75 R$_{\oplus}$ with a 1~bar atmosphere predominantly
composed of 90\% H$_2$ and 10\% N$_2$ by volume, and semi-major axes
compatible with habitable surface temperatures. Although deviations
from our fiducial model will yield different spectral features,
atmospheric concentrations, etc., the main findings summarized here
will still hold.  

Our major finding is that for H$_2$-rich atmospheres, low-UV
  radiation environments are more favorable for biosignature gas
  accumulation than high-UV radiation environments. Specifically, H is
  the dominant reactive species generated by photochemistry in an
  H$_2$-rich atmosphere. In atmospheres with high levels of CO$_2$
  atomic O will be the dominant destructive species for some
  molecules.  The low UV environments of UV quiet M stars are
favorable for accumulation of biosignature gases in an H$_2$-dominated
atmosphere. The high UV environment of sun-like and active M dwarf
stars largely prevents biosignature gas accumulation due to rapid
photochemical destruction via H  (or sometimes O), its
concentration controlled by UV photolysis.  High UV radiation is also
unfavorable for the accumulation of biosignature gases in oxidized
atmospheres \citep{segu2005}, although in contrast OH is the main
reactive species in oxidized atmospheres.

We investigated the plausibility of a number of biosignature
gases, including H$_2$,CH$_4$, H$_2$S, DMS, NH$_3$, N$_2$O, NO,
CH$_3$Cl, HCl.  While not exhaustive, we came up with some plausible
biosignature gases and others that are unsuitable as
biosignature gases, as follows.

Our list of plausible biosignature gases is dominated by Type III
biosignature gases in low-UV environments. These include CH$_3$Cl,
DMS, and N$_2$O.  Type III are gases produced for specialized
functions and therefore could well include small molecules as yet
unknown. We therefore support the idea of searching for high
concentrations of gases that do not belong in chemical equilibrium.

We also presented a new biosignature gas candidate, NH$_3$, the only
one we found reasonable as a Type I biosignature gas candidate, and
one unique to a hydrogen-rich environment. Type I are gas produced as
byproducts from energy extraction from the environment.

Our list of unlikely biosignature gases is dominated by Type I
biosignature gases, as any biosignature gases produced from energy
extraction (such as CH$_4$ or H$_2$S and numerous others), will be
either be produced by geochemical or photochemical processes or likely
rapidly destroyed by hydrogenation in a hydrogen-dominated
environment.

We have not identified any unique biosignature gas produced by any
type of photosynthesis in a thin H$_2$-rich atmosphere comparable to O$_2$
in oxidized atmospheres. In an H$_2$-dominated environment the most
likely photosynthetic byproduct is molecular hydrogen, already
prevalent in the H$_2$-dominated atmosphere, or non-volatile mineral
products. This is in contrast to O$_2$ produced by photosynthsis in
oxidized environments that is quite robust to most false positive
scenarios. (We call biosignature gases from biomass building Type II.)

Bioindicators would be helpful, but aren't easily or uniquely detectable.
The examples we gave were the hydrogen halides.

Overall, the promise of biosignature gases in H$_2$ atmospheres is
real.  We have aimed to provide a conceptual and quantitative
framework to show that there are at least some viable biosignature
gases that could be produced either by life's capture of
environmental chemical energy or are in a category of gases produced
by terrestrial life. We intend for the results here to fuel the motivation for
discovery of habitable Earths and super Earths orbiting M dwarf stars and
their atmospheric followup with the JWST.
 
\acknowledgements{We thank Jean-Michel Desert and Kartik Sheth for
  motivating questions. We thank Foundational Questions Institute
  (FQXI) for funding the seeds of this work many years ago.}

\bibliography{planets}

\end{document}